\documentclass[aps, reprint, superscriptaddress,floatfix]{revtex4-2}
\usepackage{amsmath}
\usepackage{bm}
\usepackage{graphicx}
\usepackage[caption=false]{subfig}
\usepackage{lmodern}
\usepackage{amsmath}
\usepackage{physics}
\usepackage{natbib}
\usepackage{tikz}
\usetikzlibrary{backgrounds,decorations.markings, calc, arrows.meta}
\usetikzlibrary{decorations.pathmorphing, decorations.pathreplacing}

\tikzset{>=Latex}
\tikzset{->-/.style={decoration={
  markings,
  mark=at position 0.6 with {\arrow{>[scale=0.9]}}},postaction={decorate}}
}

\tikzset{-->-/.style={decoration={
  markings,
  mark=at position 0.7 with {\arrow{>[scale=0.9]}}},postaction={decorate}}
}
\usepackage{color}
\usepackage{diagbox}
\usepackage{bm}
\usepackage{amssymb}
\usepackage{tabularx}
\usepackage{hyperref}
\usepackage{bbold}
\usepackage{xcolor}
\usepackage[nameinlink]{cleveref}
\usepackage[normalem]{ulem} 
\crefformat{section}{\S#2#1#3} 
\crefformat{subsection}{\S#2#1#3}
\crefformat{subsubsection}{\S#2#1#3}

\hypersetup{
    pdftoolbar=true,        
    pdfmenubar=true,        
    pdffitwindow=false,     
    pdfstartview={FitH},    
    pdftitle={My title},    
    pdfauthor={Author},     
    pdfsubject={Subject},   
    pdfcreator={Creator},   
    pdfproducer={Producer}, 
    pdfkeywords={keyword1} {key2} {key3}, 
    pdfnewwindow=true,      
    colorlinks=true,       
    linkcolor=carmine,          
    citecolor=carmine,        
}

\renewcommand{\(}{\left(}
\renewcommand{\)}{\right)}
\renewcommand{\[}{\left[}
\renewcommand{\]}{\right]}

\newcommand{\mc}{\mathcal}

\definecolor{vermillion}{rgb}{0.86, 0.18, 0.01}
\definecolor{AJ}{rgb}{0.0, 0.48, 0.65}
\definecolor{carmine}{rgb}{0.59, 0.0, 0.09}
\definecolor{d_blue}{cmyk}{0.91, 0.79, 0.00, 0.22}
\definecolor{go_green}{rgb}{0.13, 0.55, 0.13}

\begin{document}

\title{\texorpdfstring{Higher-order topological superconductivity in monolayer WTe$_2$ from repulsive interactions}{}}

\author{Ammar Jahin}
\affiliation{Department of Physics, University of Florida, 2001 Museum Rd, Gainesville, FL 32611, USA}

\author{Yuxuan Wang}
\affiliation{Department of Physics, University of Florida, 2001 Museum Rd, Gainesville, FL 32611, USA}

\date{\today}
\begin{abstract}
    Superconductivity has been experimentally observed in monolayer WTe2, {which in-{plane} field measurements suggested are of spin-triplet nature.}
    Furthermore, it has been proposed that with a $p$-wave pairing, the material is a second-order topological superconductor with a pair of Majorana zero modes at the corners of a finite sample. 
    We show that for {a repulsive on-site interaction and sizable Fermi surfaces,} the desired $p$-wave state arises naturally due to the Kohn-Luttinger mechanism{, and indeed a finite superconducting sample hosts corner Majorana zero modes.}
    We study the behavior of the critical temperature in response to external in-plane magnetic fields. 
    We find an enhancement to the critical temperature that depends on the direction of the magnetic field, which can be directly verified experimentally. 
\end{abstract}

\maketitle

\tableofcontents
\section{Introduction}
The understanding of the role of symmetry and topology on the behavior of materials has been one of the major ongoing developments in the field of condensed matter physics. 
Within band theory, it is now well-known that distinct phases of topological insulators and superconductors exist~\cite{Hasan_Kane_2010,Qi_Zhang_2011,Chiu_2016,Witten_2016}.
A hallmark of these topological phases is the existence of gapless excitations of the surface of a finite sample. 
For instance, the quantum spin hall (QSH) state is a 2d bulk insulating state with two counter-propagating chiral modes that are related by time-reversal symmetry. 
The QSH state has been observed in mercury telluride quantum wells ~\cite{Bernevig_2006,Markus_2007,Nowack_2013,Deacon_2017}, and, more relevant to our work, in monolayer WTe$_2$ at temperatures as high as $100~\rm K$~\cite{Yanmeng_2019,Sanfeng_2018,Tang_2017,Fei_2017,Jia_2017,Peng_2017,Akhmerov_2019}. 

{Topological superconductors are known to host gapless Majorana modes at the edges, and Majorana zero modes (MZM) bound at vortex cores. The MZMs are of special interest for their application in quantum computation~\cite{Sarma_2015,Beenakker_2013}. 
Despite intensive research and various proposed superconducting materials, unambiguous evidence for topological superconductivity with propagating Majorana modes or MZMs remains elusive. One of the key challenges is that, unlike topological insulators, the formation of intrinsic topological superconductivity requires synergy between normal state band structure and unconventional pairing symmetry induced by interaction effects.}


Recently, the concept of topological insulators and superconductors has been extended to include phases with edges that are gapped except for {``higher order boundaries", that is,} a lower dimensional locus of points on the edge. 
These new phases have been dubbed higher-order topological insulators and superconductors. 
Specifically, a nontrivial $n$-th order topology of a $d$-dimensional bulk, entails that the gapless parts of edge are $(d-n)$-dimensional~\cite{Yan_2019,Benalcazar_2017,Trifunovic_2019,Geier_2018,Schindler_2018,Wang_2018,Jahin_2022, Tiwari_2020,Langbehn_2017,Zhang_2019}. 
For example, in 2d a second-order topological superconductor will host Majorana zero modes (MZMs) on the corners of a square sample. 

Monolayer WTe$_2$ {with odd-parity pairing order} has been proposed to be one example of such higher-order topological superconductors~\cite{Hsu_2020}. 
{The material is a quantum spin Hall insulator in the normal state, and upon electric gating} has been experimentally observed to turn superconducting with a $T_c$ of about $1~\rm K$~\cite{Ebrahim_2018,Valla_2018}. 
{Interestingly, $T_c$ was found to increase in the presence of small in-plane magnetic fields, lending support to possible odd-parity pairing symmetry.}
Theoretically, in Ref.~\cite{Hsu_2020} {the authors assumed a next-nearest neighbor attractive interaction between the electrons and showed that it leads to $p$-wave pairing symmetry. 
However, the microscopic origin of the attractive interaction remains to be elucidated.}
Given the small electron density at which superconductivity is observed, in Ref.~\cite{Fu_2022} it was proposed that through virtual inter-band excitonic processes the repulsive interactions between electrons can give rise to a $p$-wave pairing instability. 
However, these results were based on {a generic band structure rather than that specific to the monolayer WTe$_2$. }



In this work, we argue that a $p$-wave superconducting order parameter in WTe$_2$ follows naturally from the Kohn-Luttinger mechanism {of electrons near the Fermi surface. 
Of course, this mechanism neglects any interband processes, which may be important at low densities, and in this sense is complementary to the pairing mechanism studied in Refs.~\cite{Fu_2022,Francisco_2022}.}
Kohn and Luttinger first showed that an isotropic electron gas in 3D will develop a superconducting instability for channels with large angular momentum under the effects of screened Coulomb interactions~\cite{KL_1965, Luttinger_1966,Layzer_1968,Chubukov_1989,Chubukov_1993,Kabanov_2011,Kivelson_2012}. 
This effect was understood as a consequence of Friedel oscillations~\cite{Friedel_1954} {that modifies the screened Coulomb interaction}, where electrons can take advantage of the attractive portions of the oscillatory potential to form a pair. 
Kohn-Luttinger type superconductivity has been recently studied for different lattice systems with short range interactions~\cite{Kivelson_2012,Rahul_2014,Thomale_2012,Thomale_2013,Raghu_2010,Vladimir_2014,Oskar_2014,Bruno_2012,Guinea_2013,Stephan_2022,Rahul_2018,Schwemmer_2022,Wolf_2018,Sondhi_2017,Korovushkin_2016,Korovushkin_2016_1,Korovushkin_2015,Korovushkin_2015_1,Stauber_2019,Srinivas_2013}. 
Even if lattice systems do not have pairing channels with arbitrarily large angular momenta, with proper geometry of the Fermi surface, superconductivity with unconventional pairing symmetry can still emerge out of repulsive interactions. 


We model the screened Coulomb interaction between electrons by an on-site repulsive interaction $U_0$ and study the resulting {pairing instability.}
At the tree-level, Cooper pairs with a non-zero amplitude of being on the same site will experience repulsion of order $U_0$ and {cannot form. }
On the other hand, pairing channels with zero {on-site components by construction, which we focus on in this work,} do not experience repulsion and have a vanishing tree-level contribution. 
The fate of these channels can only be determined by renormalizing the interaction vertex to second order in $U_0$. 

{Compared with previous works~\cite{Kivelson_2012,Rahul_2014,Thomale_2012,Thomale_2013,Raghu_2010,Vladimir_2014,Oskar_2014,Bruno_2012,Guinea_2013,Stephan_2022,Rahul_2018,Schwemmer_2022,Wolf_2018,Sondhi_2017,Korovushkin_2016,Korovushkin_2016_1,Korovushkin_2015,Korovushkin_2015_1,Stauber_2019,Srinivas_2013} on the Kohn-Luttinger superconductivity for lattice systems, }an important aspect of monolayer WTe$_2$ is the spin-orbit coupling, in which spin-singlet and spin-triplet are in general mixed.  As it turned out, despite the existence of spin-orbit coupling  in WTe$_2$~\cite{Jiang_2015, Qian_2014,Ok_2019,Choe_2016,Zheng_2016}, a spin axis is still conserved~\cite{Ok_2019, zhao_2021}, and there is an approximate residual $U(1)$ spin rotation symmetry. 
To their advantage, odd-parity electron pairs with their spin aligned in this direction necessarily have no on-site components due to the Pauli exclusion principle and evade the onsite repulsive interaction $U_0$, while all other pairing channels have on-site components.
We focus on these equal-spin channels and find that when renormalizing the interaction vertex {at one-loop order they lead to the odd-parity pairing instability} of WTe$_2$. 

Inspired by experimental observations of enhanced $T_c$ by an in-plane magnetic field, we investigate this effect using Landau-Ginzburg free energy. This was also studied in Refs.~\cite{Fu_2022,Francisco_2022}. 
However, in this work, we make sure that the term added to the free energy respects the $U(1)$ spin rotational symmetry that is present even when spin-orbit coupling is taken into account.
Furthermore, we also take into account that the direction of the spin axis and the direction perpendicular to the plane are not parallel to each other. 
We find an enhancement to $T_c$ that depends on the direction of the in-plane field. 
A magnetic field in the $\bm b$ direction yields higher enhancement to $T_c$ than a field in the $\bm a$ direction of the same magnitude. 
The reason for this direction dependence is that the spin axis defined by spin-orbit coupling is not perpendicular to the plane of the sample.   
Whereas the results of Refs.~\cite{Fu_2022,Francisco_2022} also show an enhancement to $T_c$ with a small magnetic field, the details on how the enhancement depends on the direction of the field are different from what we find, which can be directly tested experimentally.

With the odd-parity ($p$-wave) order parameter, monolayer WTe$_2$ hosts two time-reversal Kramer's partners MZMs at the same corner. 
This situation is very similar to the one studied in Ref.~\cite{Lapa_2021} of a heterostructure between a $d$-wave high-$T_c$ superconductor and a quantum spin-Hall insulator. 
There, a protocol was developed to braid the two MZMs using an in-plane magnetic field. 
Unfortunately, we show that this protocol cannot be used in the case of WTe$_2$, since the $p$-wave pairing order {condenses spin} and breaks the $U(1)$ spin rotational symmetry. 

The rest of this paper is organized as follows. 
In Sec.~\ref{sec:wte2_intro} we introduce the lattice model for WTe$_2$ in the presence of an on-site repulsive interaction.
In Sec.~\ref{sec:gap_eq} we study the {pairing} gap equation {with interactions up to one-loop accuracy, both} in real and momentum space. 
In Sec.~\ref{sec:ginz_land} we discuss the effects of an in-plane magnetic field in the sample through the formalism of a Ginzburg-Landau theory. 

\section{\texorpdfstring{Normal state of monolayer $\textrm{WTe}_2$ }{}} \label{sec:wte2_intro}
\begin{figure}[t]
    \centering
    \includegraphics{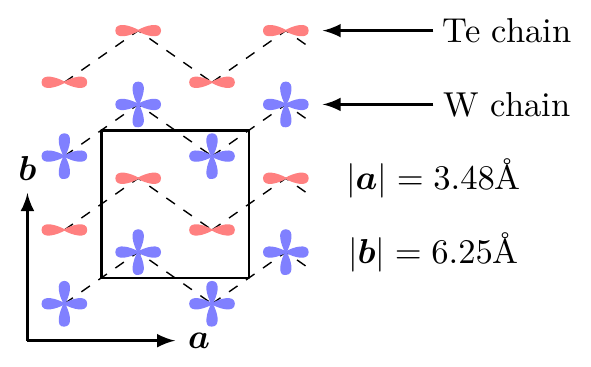}
    \caption{Monolayer WTe$_2$ unit cell. Depicted are the two W atoms and two Te atoms relevant to the low energy theory. Each W atom has a $d$ orbital, and each Te atom has a $p$ orbital.}
    \label{fig:wte_unicell}
\end{figure} 

Monolayer WTe$_2$ unit cell has two W atoms and two Te atoms that are relevant to the low energy theory~\cite{Ok_2019,Muechler_2016}. 
The lattice consists of W chains and Te chains in an alternating fashion as shown in Fig.~\ref{fig:wte_unicell}. 
The chains are parallel to the $\bm a$ direction and stacked along the $\bm b$ direction with $ \bm b \perp \bm a$.
Each W atom has a $d$-wave orbital, while each Te atom has a $p$-wave orbital.
The four orbitals per unit cell are labeled by two pseudospin degrees of freedom, $l$ and $\sigma$, where $l_z =  \pm 1$ specifies whether it is a W or a Te orbital, and $\sigma_z = \pm 1$ distinguishes their corresponding sublattice. 
Each orbital can be occupied by either a spin-up or a spin-down electron with $s_z=\pm 1/2$.

The material has an inversion symmetry around the center of the unit cell 
\begin{align}
\mc I = \sigma_x l_z.
\end{align}
Time-reversal symmetry, $\mc T = is_y K$, is also present with $K$ being the complex conjugate operation. 
Additionally, the system has a nonsymmorphic glide mirror symmetry $\mc M_x(\bm k) = s_x \[ \sigma_0 (1+e^{ik_x}) + \sigma_z (1-e^{ik_x}) \]l_z/2$ that is a combination of reflection $\bm x \to -\bm x$ and a half lattice translation in the same direction. 
Combining inversion and mirror glide gives a screw symmetry $\mc C_{2x}(\bm k) = \mc M_x(\bm k) \mc I = s_x \[\sigma_x (1+e^{ik_x}) +i \sigma_y (1-e^{ik_x})\]/2$. 
This screw symmetry is in effect a $\pi$ rotation about the $\bm a$ direction and a half lattice translation along the same direction. 

Spin-orbit coupling in WTe$_2$ is significant~\cite{Jiang_2015}, breaking the SU(2) spin rotation symmetry.
However, it was shown~\cite{Ok_2019} that, to leading order in $\bm k$, the system still has a residual spin axis and a U(1) spin rotation symmetry. We denote this spin axis as $\bm z$, and the spin-orbit coupling can be written as
\begin{align}\label{eq:no_k_soc_terms}
    H_{\text{so}} = V_{\text{so}}\ s_z \sigma_z l_y.
\end{align}
This $z$-direction is not necessarily perpendicular to the plane of the monolayer WTe$_2$, which we henceforth refer to as the $\bm c$-direction. 
Experimental measurements of the spin axis in Ref.~\cite{zhao_2021} show that the angle between the $z$ and $\bm c$ directions is $\phi_{\text{so}} = 40^\circ \pm 2^\circ$.
It is straightforward to check that such a spin-orbit coupling term is consistent with time-reversal, inversion, and screw symmetries~\cite{Ok_2019} mentioned above. Following this, we exclude additional spin-orbit coupling terms that further break the U(1) spin rotation symmetry. In what follows, we will simply refer to the $z$-component of the electronic spin as ``spin".


\begin{figure}[t]
    \centering
    \subfloat[]{\includegraphics[trim = 240 0 10 0, clip, scale=0.51]{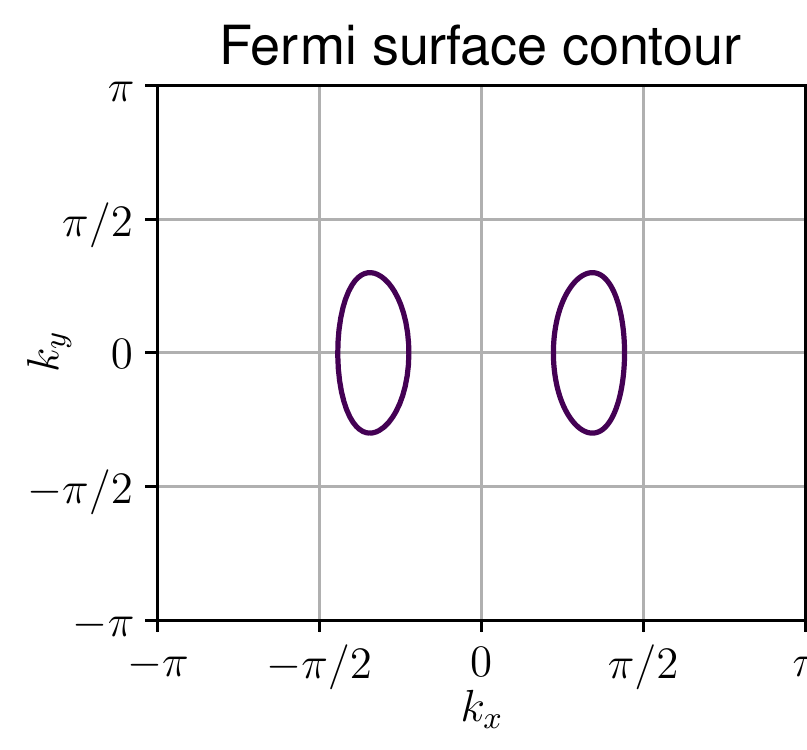}}
    \subfloat[]{\includegraphics[trim = 0 0 250 0, clip, scale=0.51]{figures/dirac_fs_cont.pdf}}

    \subfloat[]{\includegraphics[scale=0.6]{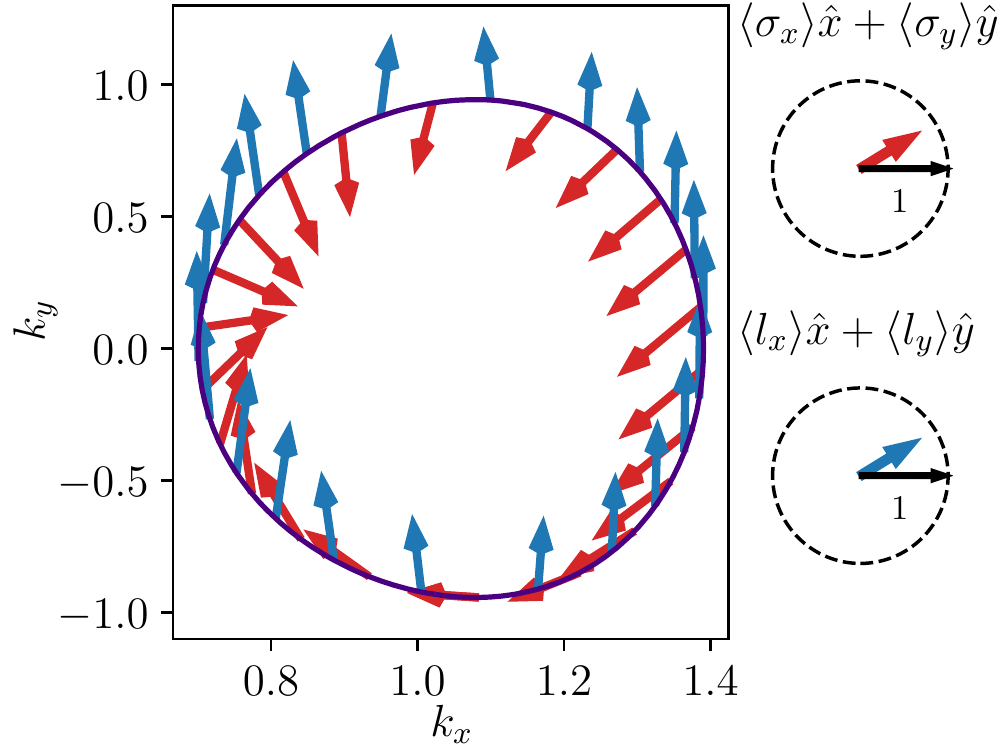}}
    \caption{Spectrum of monolayer WTe$_2$ using the tight-binding model in Ref.~\cite{Ok_2019}. Panel (a) shows the two Fermi surfaces. Panel (b) shows the gapped Dirac points. In panel (c) we plot the $\bm \sigma$ and $\bm l$ expectations values on the Fermi surface. We see that both $|\langle \bm \sigma\rangle|$ and $|\langle \bm l \rangle|$ are approximately one, indicating an approximate absence of entanglement in $\sigma$ and $l$. Further, there is a winding in $\langle \bm \sigma\rangle$, while  $\langle \bm l \rangle$  does not vary appreciably. 
    }
    \label{fig:dirac_w_fs_cont}
\end{figure}


To obtain an effective model for the low-energy fermions, we start from the tight-binding model fitted from density functional theory in Ref.~\cite{Ok_2019},
\begin{align}
\mathcal H_0(\bm k) 
 &= s_0 \otimes \begin{bmatrix} \varepsilon_d & 0 & t^{AB}_d g e^{ik_y} & t^{AB}_0 f \\ 
0 &\varepsilon_p & -t^{AB}_0 f  & t^{AB}_p g   \\
t^{AB}_d g e^{-ik_y} & -t^{AB}_0 f  & \varepsilon_d & 0  \\
t^{AB}_0 f   & t^{AB}_p g  & 0 & \varepsilon_p
\end{bmatrix} \nonumber \\ 
&+ V_{\rm so} s_z \sigma_z l_y -\mu
\end{align}
where the $2\times2$ {subblocks} represent sublattice $\sigma$ degrees of freedom. The dispersion is specified by
\begin{align}
    &\varepsilon_{d/p}(k_x) = \mu_{d/p} + 2(t_{d/p} \cos(k_x) + t'_{d/p} \cos(2k_x)), \nonumber \\
    &g(k_x) = (1+e^{-ik_x}), \qquad f(k_x) = (1 - e^{-ik_x}), 
\end{align}
and, 
\begin{align}
\mu_d = 0.4935, \ \mu_p = -1.3265, \nonumber \\
t_d = -0.28, \ t'_d = 0.075, \nonumber \\
t_p = 0.93, \ t'_p = 0.075, \nonumber \\
t^{AB}_d = 0.52, \ t^{AB}_p = 0.40, \ t^{AB}_0 = 1.02, \nonumber  \\
V_{\rm so} = 0.115,\ \mu = 0.5. 
\end{align}
We plot the Fermi-surface contours in Fig.~\ref{fig:dirac_w_fs_cont} (a). 
There are two Fermi-surface pockets each enclosing a gaped Dirac point as shown in Fig.~\ref{fig:dirac_w_fs_cont}(b). 
The gap is of order $V_{\text{so}}$. 
Moreover, in Fig.~\ref{fig:dirac_w_fs_cont}(c) we plot the expectation value of $\bm{\sigma} = (\langle\sigma_x\rangle, \langle \sigma_y \rangle, \langle\sigma_z\rangle)$ and $\bm{l}$ for the Fermi-surface in the $+ k_x$ half-plane.

We see in Fig.~\ref{fig:dirac_w_fs_cont}(c) that, to a good approximation, Bloch states on the Fermi surface are direct product states between $\sigma$ and $l$ degrees of freedom. Indeed, we see that the lengths $\langle\bm\sigma\rangle$ and $\langle\bm l\rangle$ are almost unity, indicating pure states in each sector. Moreover,  we see that $\langle\bm{l}\rangle$ is approximately a constant pointing in the $+y$ direction, while $\langle\bm{\sigma}\rangle$ approximately lies in the $xy$ plane, and winds once around the Fermi surface. This result can be understood from the  symmetry
\begin{align}
\mathcal{C}_{2x} =s_x \left [\sigma_x (1+e^{ik_x}) + i \sigma_y (1-e^{ik_x})\right ]/2
\end{align}
for all states with $k_x=0$, which in the absence of spin-orbit coupling requires the Bloch states to be polarized in $\sigma$. For small spin-orbit coupling, and for small pockets near $k_y=0$, the Bloch states are thus nearly free of entanglement in the $\sigma$ and $l$ sectors. By expanding around the Dirac points and treating other terms in the Hamiltonian as perturbations, one obtains the winding in $\langle\bm \sigma\rangle$ and polarization in $\langle\bm l\rangle$.

\begin{table}[t]
    \caption{Irreducible representations of WTe$_2$ point group. }
     \begin{ruledtabular}
         \begin{tabular}{c  c  c  c  c} 
            & \quad  $\eta_E$ \qquad  & \quad  $\eta_{\mc C}$  \qquad & \quad  $\eta_{\mc M}$ \qquad  & \quad  $\eta_{\mc I}$  \qquad \\  [0.5ex] 
          \hline 
          $A_g$ & $+1$ & $+1$ & $+1$ & $+1$ \\ 
          \hline
          $B_g$ & $+1$ & $-1$ & $-1$ & $+1$ \\
          \hline
          $A_u$ & $+1$ & $+1$ & $-1$ & $-1$ \\
          \hline
          $B_u$ & $+1$ & $-1$ & $+1$ & $-1$ \\
         \end{tabular}
     \end{ruledtabular}
 
     \label{table:irrep_point_group}
 \end{table}

With this important simplification, we write the Hamiltonian near the Dirac points as,
\begin{align}\label{eq:fs_simple_dirac}
    \mc H^{\pm}_n( \bm{\delta k}) = -( \pm v_x \delta k_x \tilde\sigma_x^{\pm} + v_y \delta k_y\tilde\sigma^{\pm}_y \pm V_{\text{so}} s_z \sigma_z )  - \mu,
\end{align}
where $H^{\pm}_n(\bm{\delta k})$ is the projected Hamiltonian with $l_y = \pm$ for the Fermi surfaces on the $\pm x$ half-plane. The Pauli operators $\tilde \sigma^\pm_\alpha$ are defined in the following way
\begin{align}
    \tilde \sigma^{\pm}_{\alpha} = e^{\pm i K \sigma_z /2} \sigma_{\alpha} e^{\mp i  K \sigma_z /2}.
\end{align}
where $\alpha=x, y$, and $K$ is the magnitude of the position of the center of any of the Fermi surfaces. This definition is necessary in order to maintain the screw symmetry of the model. See Appendix~\ref{app:hamiltoniain_screw_symm} for more details. 

While this Hamiltonian does not match the exact shape of the Fermi surfaces or the exact pseudospin texture, it does represent the correct winding and polarization in $\sigma$ and $l$ sectors and respects the symmetry.
We expect this to be sufficient to capture the symmetry of the superconducting leading instability. 

Finally, we model the electron-electron repulsion using a simple on-site interaction $U_0$. Importantly, the term ``on-site" here means that the electron density operators are not only taken at the same unit cell coordinate $\bm{R}$, but also at the same orbital and same sublattice. Of course, in the actual material, there are finite-range elements to the interaction, but we expect them to fall off rapidly as the distance grows.
Let $c_{s, \alpha}(\bm R)$ be the annihilation operator for a spin $s$ electron at the $\alpha$ orbital inside the unit cell at position $\bm R$.  Note that our notation is such that the index $\alpha=\{1,2,3,4\}$  labels both the $\sigma_z=\pm 1$ and $l_z=\pm 1$ components. 
The on-site repulsive interaction is written as
\begin{align}\label{eq:on-site_repulsive_ham}
    H_{\text{int} } = U_0 \sum_{\bm R{,\alpha}}  c^{\dagger}_{s, \alpha}(\bm R) c^{\dagger}_{s' , \alpha}(\bm R) c_{s', \alpha}(\bm R) c_{s, \alpha}(\bm R)
\end{align}
where $U_0>0$ is the strength of the interaction and repeated indices are summed over.

\section{The pairing problem}~\label{sec:gap_eq}

\begin{figure}[t]
    \includegraphics[scale=0.78]{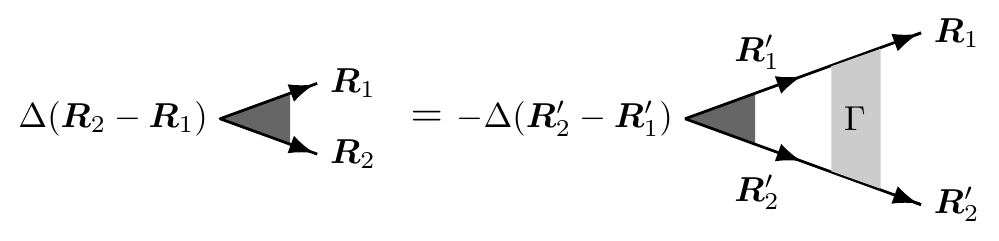}
        \caption{Linearized gap equation in real space.}
        \label{fig:linearized_gap_eqa}
\end{figure}

The point group of monolayer WTe$_2$ is $C_{2h}$, containing $E$ (the identity), $\mc C_{2x}$, $\mc M_x$, and $\mc I$, with $\mc M_x \mc C_{2x} = \mc I$, and $\mc C^2_{2x} = \mc M^2_x = \mc I^2 = E$. 
This group has four irreducible representations corresponding to the different signs of $\mc C_{2x}$ and $\mc M_x$, as shown in Table~\ref{table:irrep_point_group}.

We consider the pairing vertex $\int d\bm{k} c^\dag (\bm k) \Delta(\bm k) c^\dag(-\bm k)$, in which $\Delta(\bm k)$ is a matrix in $\sigma$, $s$, and $l$. 
All pairing wavefunctions must satisfy the Pauli exclusion principle, 
\begin{align}
    \Delta^T(-\bm k) = -\Delta(\bm k).
    \label{eq:10}
\end{align}
Solutions of the gap equation can be characterized depending on how they transform under $\mc C_{2x}$ and $\mc M_x$, 
\begin{align}\label{eq:etas_def}
    &\! \! \! \mc M_x : \Delta(\bm k) \rightarrow \mc M_x(\bm k) \Delta(\mc M_x \bm k) \mc M^T_x(-\bm k) = \eta_{\mc M} \Delta(\bm k), \\ 
    &\! \! \! \mc C_{2x} : \Delta(\bm k) \rightarrow \mc C_{2x}(\bm k) \Delta(\mc C_x \bm k) \mc C^T_{2x}(-\bm k) = \eta_{\mc C} \Delta(\bm k),
\end{align} 
where $\eta_{\mc M}$ and $\eta_{\mc C} \in \{-1, +1\}$. 
The above equations invoke the form of symmetries at both $\bm k$ and $-\bm k$ because the pairing is between electrons with opposite momenta. 
Furthermore, we refer to the $A_g$ and $B_g$ representations as $s$-wave, and the $A_u$ and $B_u$ representations as $p$-wave.

Before the projection to the vicinity of the Fermi surface, we begin by studying the pairing problem for the full lattice model in real space. 
Let $\Delta_{\alpha_1 \alpha_2}^{s_1s_2}(\bm R_2 - \bm R_1)$ be the paring amplitude between an electron at $\bm R_1, \alpha_1$ and another at $\bm R_2, \alpha_2$ with spins $s_1$ and $s_2$ respectively.  
The linearized self-consistent equation for $\Delta_{\alpha_1 \alpha_2}^{s_1s_2}(\bm R_2 - \bm R_1)$ is depicted in Fig.~\ref{fig:linearized_gap_eqa}, which is
\begin{widetext}
\begin{align}
\Delta_{\alpha_1\alpha_2}^{s_1s_2}({\bm R}_2 -\bm R_1) = -T\sum_{\omega}\Gamma^{\alpha_1 \alpha_2, \alpha'_1 \alpha'_2}(\bm R_1,\bm R_2,\bm R_1',\bm R_2') G_{\alpha_1',\alpha_1''}^{{s_1,s_1''}}(i \omega,\bm R_1' -\bm R_1'')G^{s_2,s_2''}_{\alpha_2',\alpha_2''}(-i\omega,\bm R_2' -\bm R_2'')
\Delta_{\alpha_1''\alpha_2''}^{s_1''s_2''}({\bm R}_2'' -\bm R_1''),
\label{eq:12}
\end{align}
\end{widetext}
where for compactness, repeated indices and spatial coordinates are summed over (for the rest of the paper it is generally not assumed so).
Here due to the spin rotation symmetry the Green's function $G^{s_1,s_1''}_{\alpha_1',\alpha_1''}(\omega_m,\bm R_1' -\bm R_1'')$ is diagonal in the spin sector but is a matrix in $\alpha$ and the unit cell coordinate $\bm R$. Here $\omega_m$ is a Matsubara frequency, which is summed over. 

We note that, although in general, the gap equation is difficult to solve in real space, doing so simplifies the problem for the interaction vertex $\Gamma$ at the tree level, as we shall see next.
\subsection{Tree-level pairing interaction}
\begin{figure}[t]
{\includegraphics[scale=0.85]{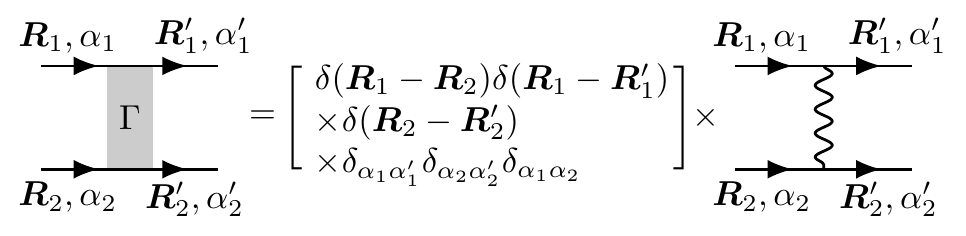}}
    \caption{Real space tree-level interaction vertex. 
    }
    \label{fig:linearized_gap_eqb}
\end{figure}

As we show in Fig.~\ref{fig:linearized_gap_eqb}, at tree-level the interaction vertex function is given by the on-site interaction
\begin{widetext}
\begin{align}
    \Gamma^{\alpha_1 \alpha_2, \alpha'_1 \alpha'_2}(\bm R_1,\bm R_2,\bm R_1',\bm R_2') =U_0\delta(\bm R_1-\bm R_2)\delta(\bm R_1-\bm R_1')\delta(\bm R_2-\bm R_2')\delta_{\alpha_1\alpha_2}\delta_{\alpha_1\alpha_1'}\delta_{\alpha_2\alpha_2'}+\mathcal{O}(U_0^2),~~~U_0>0.
    \label{eq:13}
\end{align}
\end{widetext}

As a consequence of the repulsive interaction, at leading order in $U_0$, the right-hand side of Eq.~\eqref{eq:12} is non-positive. The key insight here is that to evade the strong on-site repulsion at tree-level we expect the leading instability to be towards a pairing order that has vanishing on-site component \textit{by symmetry}; any other eigenfunction of the kernel of Eq.~\eqref{eq:12} in general has an on-site component, which invokes a negative contribution at leading order to the right-hand side of Eq.~\eqref{eq:12}. 

Naturally, equal-spin pairing orders (with Cooper pairs of electrons with the same $s_z$) have zero on-site components by symmetry due to the Pauli exclusion principle. From this argument, we conclude that the leading instability to toward equal-spin pairing. Such a pairing order experiences no repulsion at tree-level, and as we shall see, attractive pairing interactions come from $\mathcal{O}(U_0^2)$ order. Indeed, we have verified by using a momentum-space approach near the Fermi surface that all inter-spin pairing orders experience a repulsive interaction at tree-level. This is reminiscent of the fact that short-range repulsion promotes finite-angular-momentum pairing in the continuum~\cite{Luttinger_1966}.


For a multi-orbital system, in principle it is possible to have equal-spin and even-parity pairing order, e.g., by considering pairing between a W orbital and a Te orbital. 
 However, from energetic considerations of the Cooper instability, it is sensible to limit to pairing between electrons that are inversion partners. In this case, from Fermi statistics, the equal-spin pairing function is necessarily odd in parity, i.e., $p$-wave. 
 We will see this explicitly by going to $\mathcal{O}(U_0^2)$ terms in the interaction vertex.

\subsection{Renormalized interactions}
The effective attraction for the equal-spin pairing channels comes from $U_0^2$ order in the interaction vertex.
Without loss of generality, we focus only on the spin-$\uparrow \uparrow$ channel. As neither the Green's function nor the density-density interaction flips the spin, we suppress the spin indices throughout. The pairing vertex for the spin-$\downarrow \downarrow$ channel is related by time-reversal symmetry.


\begin{figure*}[t]
    \subfloat[]{\includegraphics[scale=0.6, trim = 0 -25 0 0, clip]{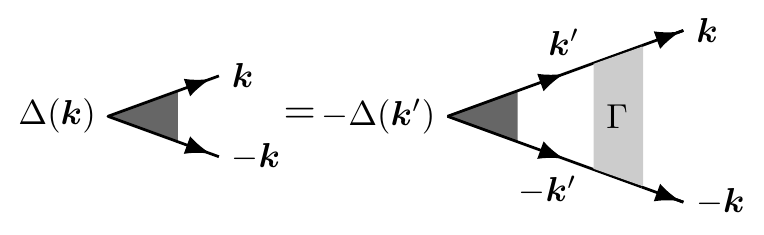}}
    \hfill
    \subfloat[]{\includegraphics[scale=0.7]{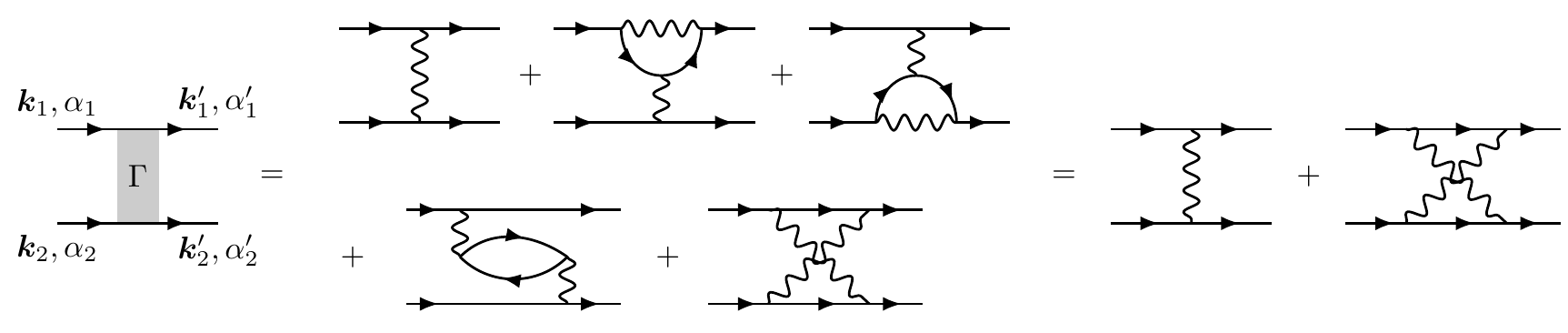}}
    \caption{(a)  Linearized gap equation in momentum space. 
        (b) Equal-spin interaction vertex to second order. 
    $\alpha_1, \alpha_2, \alpha'_1, \ \text{and, } \alpha'_2$ are combined indices for $\sigma$ and $ l$. 
    }
    \label{fig:vertex_sec_order}
\end{figure*}
In Fig.~\ref{fig:vertex_sec_order} we show the diagrams involved in the vertex calculation up to second-order terms. As is typical for Kohn-Luttinger-type pairing {with local interactions}, it is straightforward to see that the first three $\mathcal{O}(U_0^2)$ diagrams in Fig.~\ref{fig:vertex_sec_order}(b) cancel each other.
Evaluation of the last  diagrams leads to,
\begin{widetext}
{\begin{align}
\Gamma^{\alpha_1,\alpha_2,\alpha_1',\alpha_2'}(\bm k_1, \bm k_2, \bm k_1', \bm k_2') = \[U_0\delta_{\alpha_1 \alpha'_1} \delta_{\alpha_2, \alpha'_2} \delta_{\alpha_1, \alpha_2} +U_0
^2\delta_{\alpha_1 \alpha'_2} \delta_{\alpha_2 \alpha'_1} \Pi^{\alpha_1 \alpha_2}(\bm k_1 {-} \bm k_2')\]\delta(\bm{k}_1+\bm{k}_2-\bm k_1' - \bm k_2')
\end{align}}
\end{widetext}
with $\Pi^{\alpha_2 \alpha_1}(\bm q)$ being the particle-hole bubble,
\begin{align}\label{eq:pi_alpha}
    \Pi^{\alpha_2 \alpha_1}(\bm q) = - \int \frac{d\omega d \bm p}{(2\pi)^3} G^{\alpha_2 \alpha_1} (i\omega, \bm p) G^{\alpha_1 \alpha_2}(i\omega, \bm p - \bm q). 
\end{align}
dand $G^{\alpha_1,\alpha_2}(i\omega, \bm k)$ is the Green's function.

The dominant contribution to the Green's function comes near the Fermi surface. As a good approximation, we take
\begin{align}
    G(i\omega, \bm k) = \frac{\ket{\psi(\bm k)}\bra{\psi(\bm k)}}{i\omega - \varepsilon(\bm k)},
\end{align} 
{where $\ket{\psi(\bm k)}$ is the Bloch state on the Fermi surface in $\sigma$ and $l$ sectors. }
Note that because we are only considering one spin species, there is only one $\ket{\psi(\bm k)}$ at each point on the Fermi surface. 


The projected gap equation onto the Fermi surface reads,
\begin{align}\label{eq:projected_gap_eq}
    \tilde \Delta(\bm k) = - \int \frac{d\bm k'}{(2\pi)^2} \tilde \Gamma(\bm k, \bm k') \tilde \Delta(\bm k')\frac{\tanh\(\varepsilon(\bm k')/2T_c \)}{2\varepsilon(\bm k')}
\end{align}
where the projected interaction vertex $\tilde \Gamma(\bm k', \bm k) = \matrixelement*{\Psi(\bm k')}{\Gamma}{\Psi(\bm k)}$, with $\ket{\Psi(\bm k)} = \ket{\psi(\bm k)}\otimes\ket{\psi(-\bm k)}$, is given by
\begin{align}\label{eq:FS_scatter_general}
    &\tilde \Gamma(\bm k', \bm k) = U_0\sum_\alpha \psi^*_\alpha (\bm k') \psi^*_\alpha (-\bm k')  \psi_\alpha (\bm k) \psi_\alpha (-\bm k)      \\ 
    &+U_0^2  \sum_{\alpha_2 \alpha_1}  \Pi^{\alpha_2 \alpha_1}(\bm k + \bm k')  \psi^*_{\alpha_2} (\bm k') \psi^*_{\alpha_1} (-\bm k')\psi_{\alpha_1} (\bm k) \psi_{\alpha_2} (-\bm k)   \nonumber
\end{align}
and the pairing gap
\begin{align}\label{eq:project_delta}
\tilde \Delta(\bm k) = \sum_{\alpha \beta} \psi^*_\alpha(\bm k) \psi^*_\beta(-\bm k) \Delta_{\alpha \beta}(\bm k)
\end{align} 
is the projection of $\Delta(\bm k)$ onto the Fermi-surface. 
{Plugging Eq.~\eqref{eq:10} into Eq.~\eqref{eq:project_delta} we obtain that by Fermi statistics $\tilde \Delta(\bm k) = -\tilde \Delta(- \bm k) $.
Since the Fermi surface states are non-degenerate for each spin, inversion simply takes $\tilde \Delta(\bm k)$ to $\tilde \Delta(-\bm k)$, and the equal-spin pairing gap corresponds to a $p$-wave order.}

\begin{figure}[t]
    \includegraphics[scale=0.8]{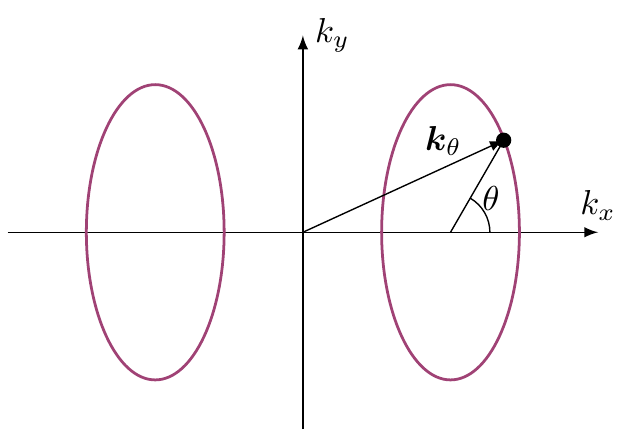}
    \caption{Definition of $\bm k_{\theta}$ on the Fermi surface.}
    \label{fig:theta_def}
\end{figure}

The integral over energy in Eq.~(\ref{eq:projected_gap_eq}) gives the usual $\log(\Lambda/T_c)$, where $\Lambda$ is the energy cut-off.
What is left is the angular integral over the oval Fermi surfaces.
For the $p$-wave solution we can rewrite the self-consistent equation as,
\begin{align}
    \Delta(\theta) = -\int d\theta'  \ \Gamma(\theta, \theta') \nu(\theta') \Delta(\theta')   \log(\frac{\Lambda}{T_c}) 
\end{align} 
where $\theta$ parametrizes the momentum on the Fermi-surface as shown in Fig.~\ref{fig:theta_def}, $\Delta(\theta) = \Delta(\bm k_\theta)$, and
\begin{align}
    \Gamma(\theta, \theta') = \tilde \Gamma(\bm k_\theta, \bm k_{\theta'}) - \tilde \Gamma(\bm k_\theta, -\bm k_{\theta'}).
\end{align} 
We use the density of states $\nu(\theta)$ for the model in Eq.~(\ref{eq:fs_simple_dirac})
\begin{align}
    \nu(\theta) = \frac{1}{(2\pi)^2} \frac{\mu}{{v^2_x \cos^2(\theta) + v^2_y\sin^2(\theta)}}. 
\end{align}
{The pairing gap is an eigenfunction given by}
\begin{align}\label{eq:theta_eigenvalue_prob}
    \lambda \Delta(\theta) = -\int d \theta' \  \Gamma(\theta, \theta') \nu(\theta') \Delta(\theta').   
\end{align}
As usual, pairing channel 
with $\lambda > 0$ has a critical temperature 
\begin{align}
    T_c =\Lambda \exp\(-\frac{1}{\lambda}  \).
    \label{eq:25}
\end{align}

One feature of $\Gamma(\theta, \theta')$ that is immediately obvious is that the first-order contribution in $U_0$ is zero. 
This can be seen from Eq.~(\ref{eq:FS_scatter_general}), as terms proportional to $U_0$ are symmetric under $\bm k \rightarrow -\bm k$, and thus cancel out in $\Gamma(\theta, \theta')${; this is precisely what we discussed in the previous subsection.}
In the following, we focus on second-order terms for the scattering matrix $\Gamma$. 

We evaluate $\Gamma(\theta, \theta')$ numerically. 
In Fig.~\ref{fig:same_v_op_fs_gamma} we show our result for $\tilde\Gamma(\bm k_{\theta}, \bm k_{\theta'})$ and $\tilde\Gamma(\bm k_{\theta}, -\bm k_{\theta'})$, where $\bm k_{\theta}$ and $\pm\bm k_{\theta'}$ belong to the same/different Fermi surface.
The most pertinent feature of the result is that $\Gamma(\bm k_{\theta}, \bm k_{\theta'})< \Gamma(\bm k_{\theta}, -\bm k_{\theta'})$ for all $\theta$ and $\theta'$. 
This lead to $\Gamma(\theta, \theta') < 0$ and to $\lambda > 0$.


We note that our analysis relies on an expansion around the Fermi surface. 
For small values of $\mu$ the interplay between the conduction and valence band may become important and, as studied in Refs.~\cite{Fu_2022,Francisco_2022}, can lead to odd-parity superconductivity without consideration of the Kohn-Luttinger effects. The Kohn-Luttinger effects we consider here are in this sense complimentary to the results there, and further elucidate the exact form of the odd-parity pairing order.


\subsection{Origin of effective attraction}
\begin{figure}[t]
    \centering
    \includegraphics[scale=0.55]{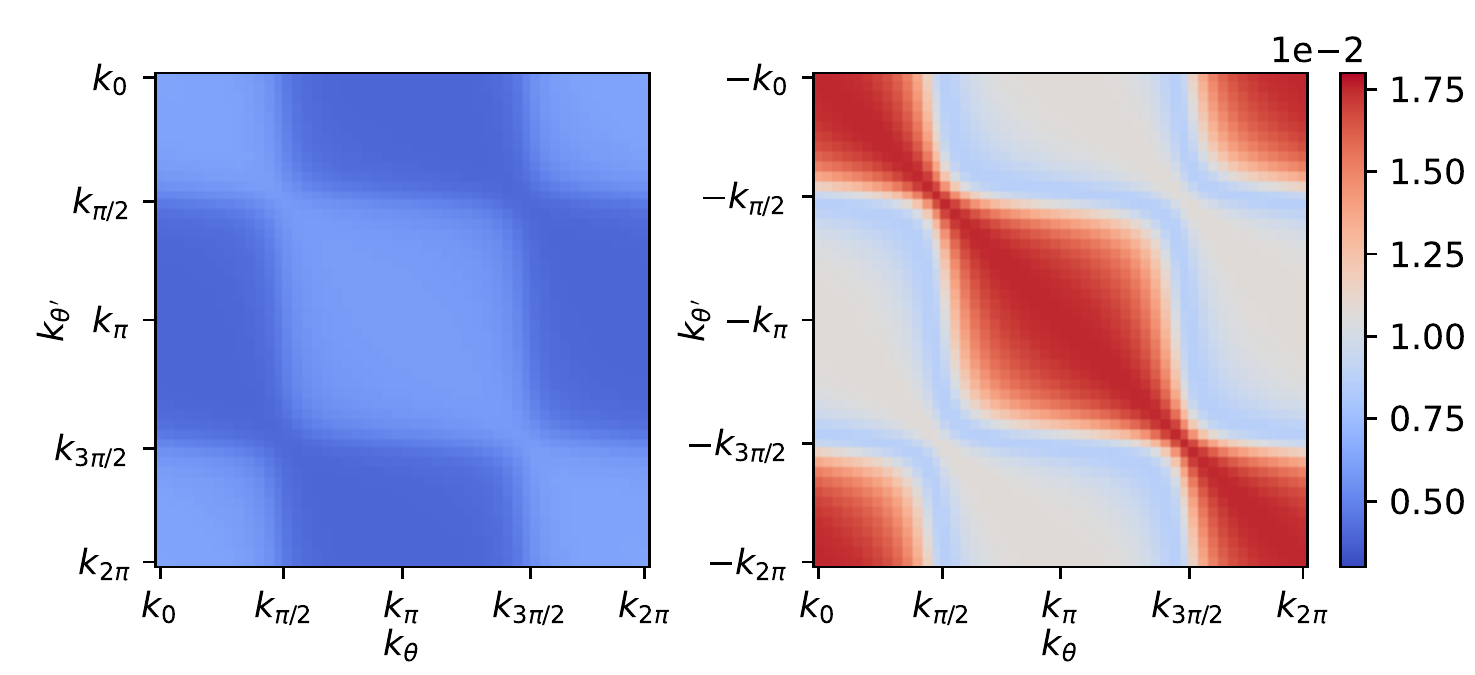}
    \caption{Numerical evaluation of the second-order terms of $\tilde{\Gamma}(\bm k_{\theta'}, \bm k_{\theta})$ (left) and $\tilde{\Gamma}(-\bm k_{\theta'}, \bm k_{\theta})$ (right). As discussed in the main text, we see that in general $\tilde{\Gamma}(\bm k_{\theta'}, \bm k_{\theta}) < \tilde{\Gamma}(-\bm k_{\theta'}, \bm k_{\theta})$ which is important in obtaining a $p$-wave superconducting channel with positive $\lambda$. $\mu = 0.3, v_x = 1, v_y=1/3, V_{sc} = 0.1$.}
    \label{fig:same_v_op_fs_gamma}
\end{figure}

The details of Fig.~\ref{fig:same_v_op_fs_gamma} are complicated by the pseudospin texture on the Fermi surfaces. 
However, the important features that make solutions with positive $\lambda$ possible are easily understood from the polarization bubble $\tilde{\Gamma}(\bm k, \bm k')$, which is
\begin{align}
    \Pi(\bm q) =
    \int \frac{d \bm p}{(2\pi)^2} \frac{f(\varepsilon(\bm p)) -f(\varepsilon(\bm p - \bm q))  }{\varepsilon(\bm p)- \varepsilon(\bm p - \bm q)},
\end{align}
where $\bm q = \bm k + \bm k'$, and $f(\varepsilon)$ is the Fermi-Dirac function. 
The integrand of $\Pi(\bm q)$ is peaked when $\bm p$ and $\bm p - \bm q$ are both near one of the Fermi surfaces. For small $\bm q$, the dominant contribution comes from when both $\bm p$ and $\bm{p-q}$ are on one of the two Fermi pockets, while for $\bm q$ larger than the size of the Fermi pocket, the dominant contribution can only come from when $\bm p$ is on one Fermi pocket and $\bm {p-q}$ is on another. We hence have, e.g., $\Pi(0) = 2 \Pi(2\bm K)$, where $2\bm K$ is the separation of the two Dirac points. {This relation also extends to that between $\Pi(\bm q) = 2 \Pi(2\bm K+\bm q)$, where $\bm q$ is smaller than the size of a Fermi pocket.} For this reason, the projected interaction vertices satisfy $\tilde{\Gamma}(\bm k_{\theta'}, \bm k_\theta) < \tilde{\Gamma}(-\bm k_{\theta'}, \bm k_\theta)$ as found in Fig.~\ref{fig:same_v_op_fs_gamma}.




{Analytically, $\Pi(0)$ is nothing but the density of states $\nu(\mu)$ at the Fermi level. For $\mu\gg V_{\rm so}$, the density of states is given by a Dirac-like spectrum with a linear density of states, and thus  $\Pi(0)\propto \mu$.
In this situation, the coupling constant in Eq.~\eqref{eq:25} satisfies $\lambda\propto \mu$.
Using this, and our discussion above, 
our result indicates an exponential increase in the critical temperature with increasing chemical potential at large dopings. At small dopings when $\mu\gtrsim V_{\rm so}$, $\Pi(0)=\nu(\mu)$ tends to a constant. However, as we discuss below, $T_c$ will be exponentially suppressed by another mechanism. Experimentally, a sharp increase in superconductivity with doping has been observed in Refs.~\cite{Ebrahim_2018, Valla_2018}, even at very small doping. Therefore additional pairing mechanism such as that from inter-band effects~\cite{Fu_2022,Francisco_2022} is likely required to explain the experimental data.}

From Eq.~\eqref{eq:FS_scatter_general}, another contribution to $\tilde\Gamma(\bm k, \bm k')$ comes from the matrix elements due to the pseudospin texture of the Fermi surfaces. 
An essential ingredient to achieving the effective attraction between the Cooper pairs is the Dirac dispersion, which provides an in-plane component to the spinors on the Fermi surface. 
{To show this, it is helpful to consider a hypothetical situation without the Dirac dispersion, in which} the spinors are completely determined by the spin-orbit coupling. 
In this case, we can take the spinors on each Fermi surface as constant. 
Focusing on the equal spin states with $+s_z$, the relevant spinors are \textit{polarized} as $ \psi(\bm k) = \ket{+s_z, +\sigma_z, +l_y}$ for the Fermi surface at $\bm K$ and $\psi(-\bm k) = \ket{+s_z, -\sigma_z, -l_y}$ for the opposite Fermi surface. 
{One can directly evaluate $\tilde \Gamma (\bm k', \bm k)$ in Eq.~(\ref{eq:FS_scatter_general}) and verify that, at both $U_0$ and $U_0^2$ orders,  it vanishes for any $\bm k$, and $\bm k'$ on either of the two Fermi surfaces. Indeed, the polarized spinors indicate that the leftmost external fermion lines in the Kohn-Luttinger diagram (the last one in Fig,~\ref{fig:vertex_sec_order})  relevant to the pairing process (with opposite momenta) correspond to fermions fully polarized on opposite sublattices (with $\sigma_z=\pm 1$).
Since the interaction is on-site, which enforces the same sublattice index for all four fermion operators, the scattering amplitudes by both of the wavy lines are zero.
By contrast, the presence of the Dirac dispersion ensures the fermions on the Fermi surface are located coherently on both sublattice sites (in the limit of no spin-orbit coupling, a Bloch state on the FS is an equal superposition between two sublattices), and thus the scattering amplitude is nonzero even for pairing between fermions with opposite sublattice composition.} {At low dopings, the sublattice texture on the FS is dominated by the spin-orbit coupling and nearly polarized. In this situation pairing from the Kohn-Luttinger mechanism is indeed exponentially suppressed, and as we mentioned additional pairing mechanism e.g., interband effects, is indeed needed.}



{Therefore, we conclude that both the Fermi surface geometry and the underlying Dirac points, even though they are gapped by the spin-orbit interaction, lead to superconductivity due to the Kohn-Luttinger mechanism in WTe$_2$. We note that the same effect was shown to lead to superconducting instabilities in a graphene-like system~\cite{Rahul_2014}. Due to the different point group symmetry, the pairing symmetry there was found to be $f$-wave. As we shall see, another difference is the role of spin-orbit coupling, which is negligible in graphene but not in the present system.}


\subsection{Order parameter symmetry representation}
Our theory only puts constraints on the projection of the order parameter on the Fermi surfaces. 
These constraints have already pinned down the order parameter to be a $p$-wave. 
However, we might still ask if the Kohn-Luttinger mechanism favors $A_u$ representations over $B_u$ representations or vice versa.

Consider the following $\Delta(\bm k)$ making an $A_u$ representation,
\begin{align}\label{eq:BdG_pairing_term_1}
    \Delta_1 (\bm k) \propto \sin(k_x) s_0 \sigma_0 l_0.
\end{align}
Being an $p$-wave equal-spin pairing channel, we only need to check if it has a non-zero projection onto the Fermi surface in order to determine if it is a valid order parameter. 
Using Eq.~(\ref{eq:project_delta}), and the states near the Fermi-surface from the simplified Hamiltonian in Eq.~(\ref{eq:fs_simple_dirac}), $\ket{\psi(\bm k)} =  (1,\ \pm i)^T \otimes (\cos(\theta /2) , \ \sin(\theta/2) e^{i\phi} )^T/ 2 $, and $\ket{\psi(-\bm k)} =  (1,\ -(\pm i))^T \otimes (\sin(\theta /2), \ \cos(\theta/2) e^{-i\phi} )^T/ 2 $, we find the projection of $\Delta_1(\bm k)$ onto the Fermi surface
\begin{align}
    \tilde \Delta_1(\bm k) &\propto \sin(k_x) \sum_\alpha \psi^*_\alpha(\bm k) \psi^*_\alpha(-\bm k) s_0 \nonumber  \\
    &= \sin(k_x) \sin(\theta) s_0.
\end{align}
In the case of zero spin-orbit coupling, we have $\theta = \pi/2$, and as spin-orbit coupling increases $\theta$ decreases. 

Meanwhile, $B_u$ representations can also describe an equal-spin $p$-wave with non-zero projection into the Fermi surface. 
Consider the following example,
\begin{align}\label{eq:BdG_pairing_term_2}
    \Delta_2(\bm k) \propto \sin(k_x) s_z \sigma_0 l_0,
\end{align} 
which has a projection, 
\begin{align}
    \tilde \Delta_2(\bm k) \propto \sin(k_x) \sin(\theta) s_z.
\end{align}
{We see that $\Delta_1$ and $\Delta_2$ correspond to the equal and opposite pairing amplitudes among spin-up and spin-down fermions.} Due to the additional  $U(1)$ spin rotation symmetry that relate the two, both $A_u$ and $B_u$ representations are equally favored by the Kohn-Luttinger mechanism. 

It is important to note that the superconducting state $\Delta_1$ and $\Delta_2$ would break the $U(1)$ spin rotation symmetry $e^{i\theta s_z}$. This can be seen by writing 
\begin{align}\label{eq:32}
\Delta(\bm k) \propto \bm d \cdot \bm s\, is_y,
\end{align} in which the $\bm d$ vector is along $\hat y$ and $\hat x$ directions for $\Delta_1$ and $\Delta_2$  respectively, thus breaking the spin rotation symmetery in the $xy$ plane.

\begin{figure}
    \subfloat[]{\includegraphics[scale=0.41, trim = 0 0 10 0, clip]{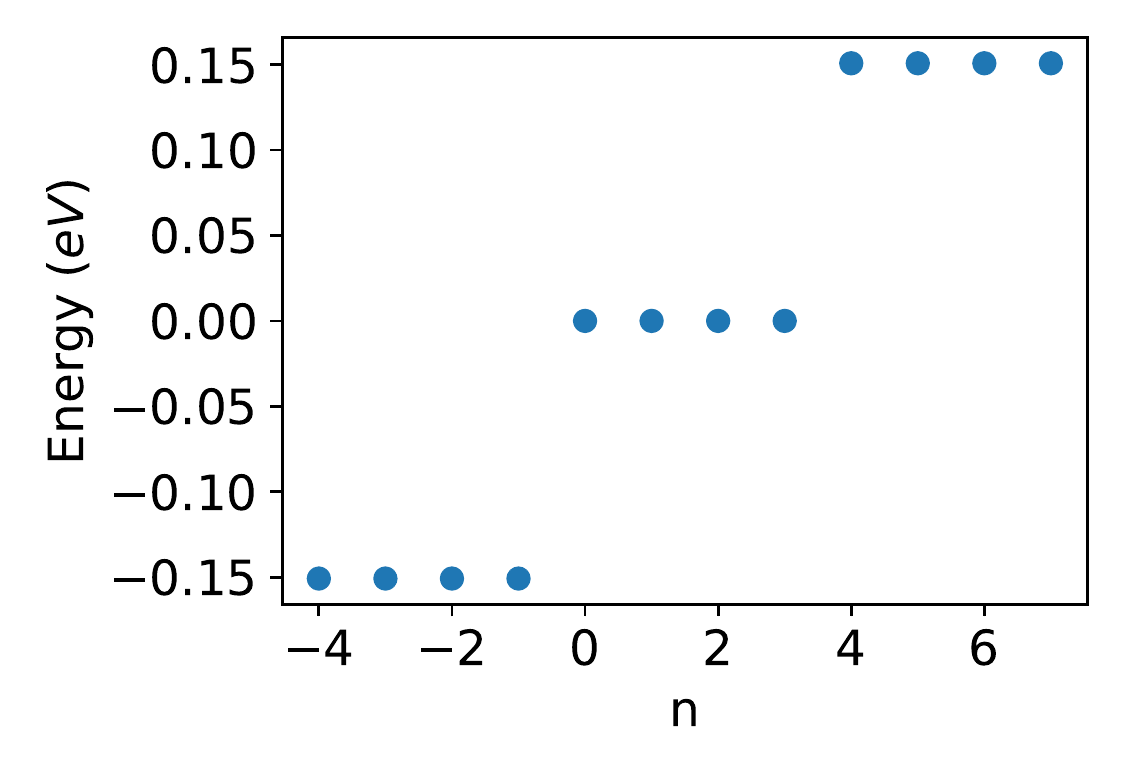}}
    \subfloat[]{\includegraphics[scale=0.425, trim = 18 0 0 0, clip]{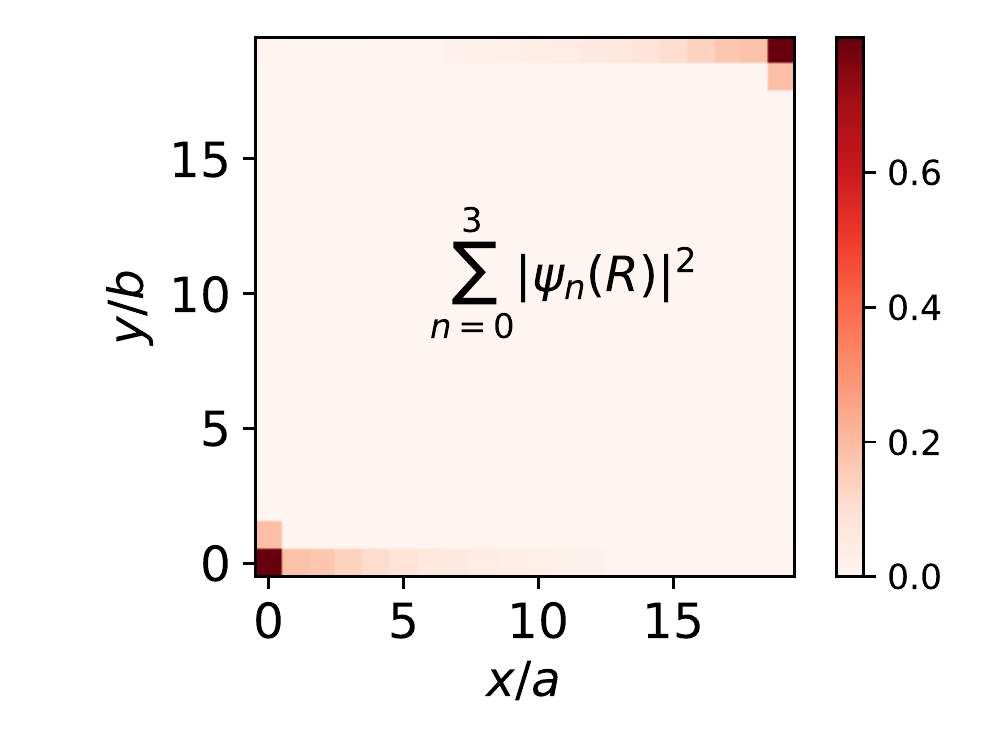}}
    \caption{Panels (a) and (b) show the existence of four Majorana Zero modes that are localized on two opposite corners of the sample. Each of the opposite corners has two Majorana zero modes that are Kramer's partners. 
    }
    \label{fig:corner_modes}
\end{figure}
\subsection{Majorana zero modes and braiding}
In this section, following the identification in Ref.~\cite{Hsu_2020}, we demonstrate the existence of Majorana corner modes in WTe$_2$ with the leading pairing instability dictated by the Kohn-Luttinger mechanism.
We start with the DFT-fitted tight-binding model in Ref.~\cite{Ok_2019} and add the pairing terms in Eqs.~(\ref{eq:BdG_pairing_term_1}) and (\ref{eq:BdG_pairing_term_2}).
In Fig.~\ref{fig:corner_modes} (a) we show the distribution eigenvalues around zero energy. 
We clearly see four modes that are pinned at zero energy. 
With time-reversal $\mc T^2 = -1$, these modes are two Kramers pairs Majorana zero modes.
In Fig.~\ref{fig:corner_modes} (b) we plot the average probability distribution in real space for the Majorana zero modes. 
One Kramer pair is localized at one corner, and the other pair is localized on the opposite corner related by inversion.  
{This is consistent with the results in Ref.~\cite{Hsu_2020}.}

Time reversal symmetry can be broken by the introduction of a magnetic field
\begin{align}
    H_{\text{mag}} = h_x s_x \sigma_0 l_0 + h_y s_y \sigma_0 l_0. 
\end{align}
With time reversal symmetry broken, the Majorana zero modes at each corner are no longer Kramer's partners, and they can hybridize and gap each other out. 
This is also confirmed by our numerical study.
This is important to contrast with the Majorana pairs discussed in Ref.~\cite{Lapa_2021}. 
There, the MZMs were obtained from a $d$-wave superconductor sitting on top of a QSH state. 
Such heterostructure also results in a Kramer's pair of MZMs at the corners. 
Using this platform, it is possible to braid the pair of Majoranas by tuning the in-plane magnetic fields. 
Magnetic fields also break time-reversal in the heterostructure, however, 
{due to the spin-singlet nature of the pairing order}
the system still has a $U(1)$ spin rotation symmetry in the superconducting state. 
Thus in the presence of an in-plane magnetic field, the original time-reversal symmetry $\mc T$ can be combined with the $U(1)$ symmetry to make a new time-reversal symmetry $\mc T' = e^{i\pi/2 s_z} \mc T$ that squares to $+1$. 
Together with the particle-hole symmetry of the BdG Hamiltonian, this puts the heterostructure in the symmetry class BDI. 
This 
{composite symmetry ensures} that a pair of MZMs of the same chiral eigenvalue is protected, and is one of the prerequisites for braiding the MZMs using in-plane fields.
Since the $p$-wave superconducting order of WTe$_2$ breaks the spin $U(1)$ symmetry, the protocol developed in Ref.~\cite{Lapa_2021} cannot be used to braid the MZMs. 
{It remains to be seen what perturbations can be added to manipulate and possibly braid the two MZM's at each corner for the intrinsic $p$-wave state. Alternatively, one can introduce a $d$-wave pairing order to monolayer WTe$_2$ via proximity effect.}

\begin{figure}
    \centering 
    \includegraphics{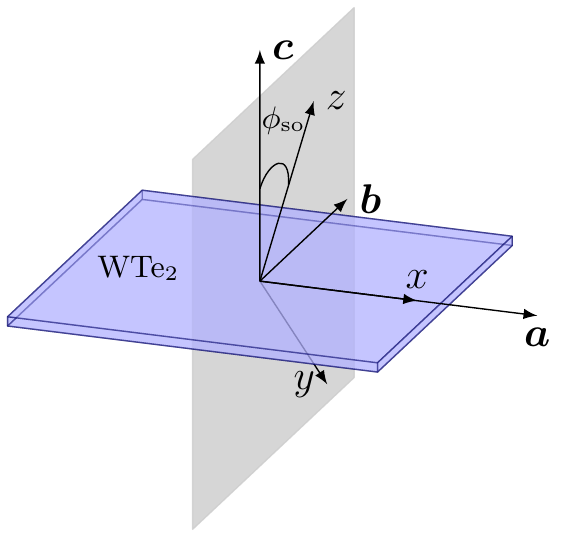}
    \caption{The $\bm a$ and $\bm b$-directions are the lattice vectors for WTe$_2$, and the $\bm c$-direction is perpendicular to the WTe$_2$ plane. The $z$-direction is the direction of the spin axis picked up by spin-orbit coupling. The $x$-direction is chosen to be parallel to the $\bm a$-direction, and the $y$-direction is perpendicular to the $xz$-plane. }
    \label{fig:different_coordinates}
\end{figure}

\section{Effects of in-plane external magnetic field}\label{sec:ginz_land}
In this section we study the effect of in-plane external magnetic fields on $T_c$, using a Ginzburg-Landau theory.  
In the presence of a magnetic field, the free energy can be written in a form that's explicitly invariant under the $U(1)$ spin rotation
\begin{align}
    F = (\alpha(T) + \chi B^2)(\bm d^* \cdot \bm d) + 2 \gamma |d_z|^2 \nonumber \\
    + \mu \bm B \cdot (i\bm d \times \bm d^*)+ \eta |\bm B \cdot \bm d|^2 
\end{align}
{where we denote the pairing order parameter by their $\bm d$ vectors defined in Eq.~\eqref{eq:32}, }$\bm$ $\alpha=\kappa (T - T_c(\bm B = 0))$, and $\kappa,\gamma,\mu,\eta,\chi > 0$. 
Here $\mu$ {(not to be confused with the chemical potential)} describes the Zeeman effect caused by the magnetic field, and the $\eta$ term accounts for pair breaking due to orbital effects. 
The $\gamma$ term is due to the effect of spin-orbit coupling. 
{Indeed, in the absence of spin-orbit coupling, the system has a full $SU(2)$ spin rotation symmetry and all three $\bm d$ vectors are equally favored by symmetry.} 
We take $\gamma > 0$ such that with no magnetic field we have the equal-spin channels $d_x$ and $d_y$ being the leading instabilities of the system, as predicted by our Kohn-Luttinger analysis.

{Note that the $z$-direction is not perpendicular to the sample plane; rather it is the direction of spin-polarization due to spin-orbit coupling~\cite{zhao_2021}. }
Thus in the following, it is important to distinguish between two different coordinate systems. 
The $a,b,c$-coordinate system is such that the $c$ direction is perpendicular to the plane of the sample, and the $a$ and $b$ directions are as defined in Fig.~\ref{fig:wte_unicell}. 
On the other hand, the $x,y,z$-coordinate system is such that $x \parallel a$, while the $z$-direction 
is in the $cb$-plane at an angle $\phi_{\text{so}}$ with the $c$-direction.  
See Fig.~\ref{fig:different_coordinates} for an illustration.

\begin{figure}
    \centering 
    \subfloat[]{
    \includegraphics[scale=0.5]{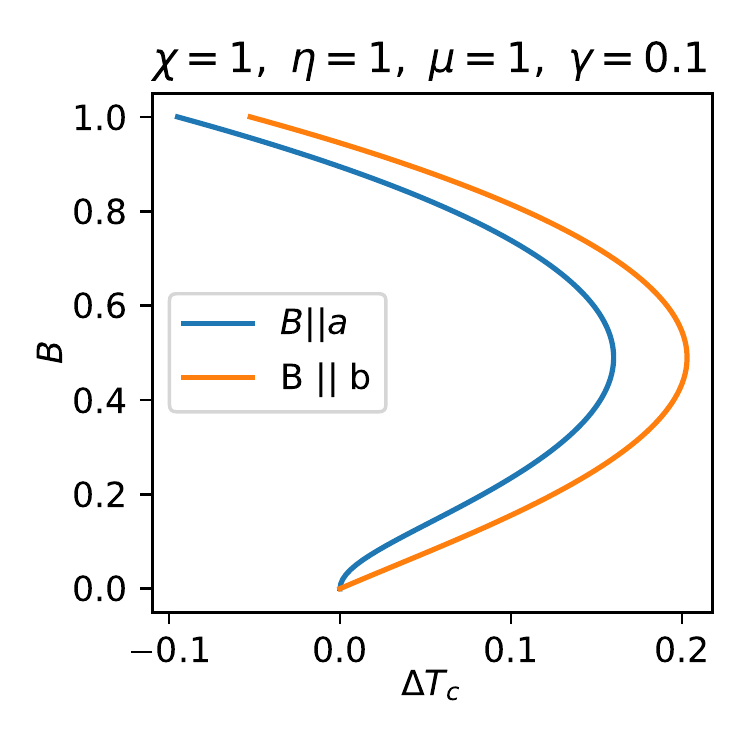}}
    \subfloat[]{
    \includegraphics[scale=0.5]{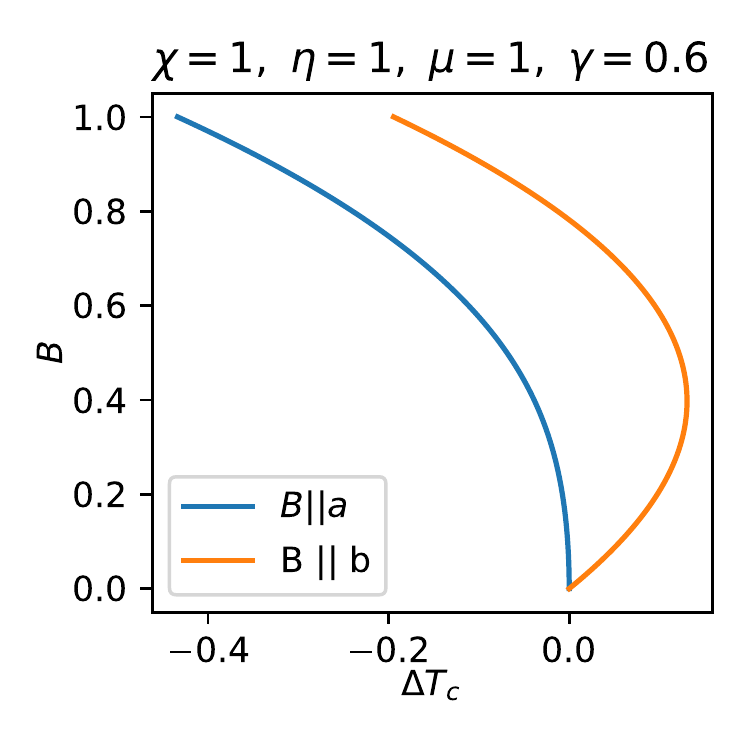}}
    \caption{The dependence of the critical temperature on in-plane magnetic fields, and the strength of the spin-orbit coupling $\gamma$. For $\gamma < \mu^2 / 2\chi$, magnetic fields in the $\bm a$  and $\bm b$-directions causes an enhancement to $T_c$, with the $\bm b$ field direction causing bigger enhancement. 
    However, for $\gamma > \mu^2 / 2\chi$, magnetic fields in the $\bm a$ direction always cause $T_c$ to decrease, while we still see enhancement for fields in the $\bm b$ direction. }
    \label{fig:critical_field}
\end{figure}

We separately study the effect of in-plane magnetic fields along $a$ and $b$ directions. 
For $\bm B = B_a \bm a$ the free energy reduces to, 
\begin{align}
    F = (\alpha(T) + \chi B_a^2)(\bm d^* \cdot \bm d) + 2 \gamma |d_z|^2 \nonumber \\
    + \mu B_a i[d_y d_z^* - d_z d_y^*] + \eta |B_a d_x|^2.
\end{align}
In this case, the $x$-component of the order parameter does not mix with the other two, and setting $d_x = 0$ {always }results in a {higher} critical temperature.
Solving the eigenvalue problem for the $y$ and $z$ components gives a critical temperature, 
\begin{align}\label{eq:tc_phi_zero}
    T_c(B_a) = T_c(0) - \chi B_a^2 - \gamma + \sqrt{\gamma^2 + \mu^2 B_a^2 }. 
\end{align}
Looking at the behavior near $B_a = 0$ we have $T_c(B_a) = T_c(0) - (\chi -  \mu^2/2\gamma)B_a^2$. For $\gamma < \mu^2/2\chi$, small fields enhance the critical temperature. 
However for $\gamma > \mu^2 / 2\chi$ any field is detrimental to superconductivity.  
{We plot both scenarios in Fig.~\ref{fig:critical_field} (shown as orange curves).}

Next we consider $\bm B = B_b \bm b$, for which the free energy can be written as, 
\begin{align}
    F = \(\alpha(T) + \chi B_b^2\)(\bm d \cdot \bm d^*) + 
    2\gamma|d_z|^2 \nonumber \\  + \mu B_b\[d_c d_a^* - d_a d_c^*\] + \eta |B_ b d_b|^2.
\end{align} 
Unlike the previous case, {because of the misalignment between $c$ and $z$,} all components of $\bm d$ are mixed{, and the behavior of the system depends on} $\phi_{\text{so}}$.
For $\phi_{\text{so}} = 0$ we have a situation very similar to that when $\bm B = B_a \bm a$, and the critical temperature reduces to Eq.~(\ref{eq:tc_phi_zero}). 
The other {limiting} case is when $\phi_{\text{so}} = 90^\circ$, such that $\hat z = \bm b$. In this case, the $z$-component of the order parameter decouples from the other components. 
The critical temperature, in this case, is, 
\begin{align}
    T_c(B) = T_c(0) - \chi B^2 + \mu B.
\end{align}
When $\phi_{\text{so}} = 90^\circ$ the critical temperature increases linearly with the magnetic field for small fields. 
This is a bigger enhancement to $T_c$ than what we get when $B \parallel \bm a$. 
Additionally, the enhancement does not depend on how big or small $\gamma$ is. 
Real WTe$_2$ however has $\phi_{\text{so}} \approx 40^\circ$. 
We thus expect the enhancement of the critical temperature with small fields to be somewhere between the two limiting cases. 

For the general case when $\bm B \parallel \bm b$ we write
\begin{align}
    F = \bm d^\dagger \[(\alpha + \chi B^2) \mathbb{1} +  M\] \bm d 
\end{align}
with 
\begin{align}
    M = \begin{bmatrix}
        0 && 0 && i\mu B \\ 
        0 && \eta B^2 + 2\gamma \sin^2(\phi_{\text{so}}) && \gamma \sin(2\phi_{\text{so}}) \\ 
        -i \mu B && \gamma \sin(2\phi_{\text{so}}) && 2\gamma \cos^2(\phi_{\text{so}})
    \end{bmatrix}
\end{align}
The problem of finding the $T_c$ reduces to the problem of finding the smallest eigenvalue of $M$. 
The characteristic equation $\det(M - \lambda \mathbb{1}) = 0$ is cubic and does not have an easy solution. 
In Fig.~\ref{fig:critical_field} we show the numerical solution to the characteristic equation for $\phi=40^\circ$. 
We compare the enhancement to $T_c$ for when $\bm B \parallel \bm a$ and for when $\bm B  \parallel \bm b$. 
Fig.~\ref{fig:critical_field} clearly shows that $\bm B  \parallel \bm b$ yields a bigger enhancement to $T_c$ than $\bm B  \parallel \bm a$. 
In fact when $\gamma > \mu^2 / 2\chi$ Fig.~\ref{fig:critical_field}(b) shows that we only get enhancement when $\bm B \parallel \bm b$ and not when $\bm B \parallel \bm a$.
{This is the key finding of our work. The field dependence of $T_c$ has been measured for monolayer WTe$_2$ with similar behaviors observed~\cite{Li_2018}. It would thus be interesting to examine the dependence of $T_c$ on the direction of the in-plane Zeeman field. }

The same problem was studied in Ref.~\cite{Fu_2022}, and we point out two ways in which our results differ. 
First, here we make sure that the spin-orbit coupling term included in the free energy respects the $U(1)$ spin rotation symmetry the normal state enjoys. 
{On the other hand, the authors of Ref.~\cite{Fu_2022} assumed $d_z$ and $d_y$ are degenerate even though they are not related by any symmetries.}
Second, we take into account the fact that the spin axis in the $z$-direction is not parallel to the $c$-direction. 
These considerations lead to qualitatively different results. 
For example, for $\gamma > 0$, results in  Ref.~\cite{Fu_2022} is that $\bm B \parallel \bm a$ will not enhance $T_c$, whereas our result shows that for a sufficiently small spin-orbit coupling, we get $T_c$ enhancement. 
Furthermore, for $\bm B \parallel \bm b$, results in Ref.~\cite{Fu_2022} is that for a sufficiently large spin-orbit coupling, there is no enhancement to $T_c$, while our results show that a small field will always lead to an enhancement in $T_c$, regardless of the value of $\gamma$. 




\section{Discussion}
In this work, we studied the superconducting ground state of monolayer WTe$_2$ from on-site repulsive interactions. 
{By constructing an effective model for the Fermi surfaces of WTe$_2$, }we show that the leading instability of the system is an equal-spin $p$-wave channel. 
This suggests that WTe$_2$ is a second-order topological superconductor with a pair of Majorana zero modes on the corners of a finite sample.

We also studied the effect of an external magnetic field on superconducting monolayer WTe$_2$. 
Using a Ginzburg-Landau theory, we study the effect of in-plane magnetic fields on $T_c$. 
We find an enhancement to $T_c$ at small field strengths. 
Interestingly, this enhancement is dependent on the direction of the in-plane field. 
In particular, a field in the $\bm b$ direction yields a bigger enhancement as compared to a field of the same strength but oriented in the $\bm a$ direction. 
This discrepancy can be used as an experimental tool to confirm the nature of the order parameter in WTe$_2$. 


\acknowledgments
We thank Andrey Chubukov, Meng Cheng, and Julian Ingham for the useful discussions. This work is supported by NSF under award number DMR-2045781.

\appendix


\section{{Screw symmetry  $C_{2x}$ in the simplified Hamiltonian}}\label{app:hamiltoniain_screw_symm}
Special care is needed when writing down the simplified Hamiltonian in Eq.~(\ref{eq:fs_simple_dirac}) in order to make sure it respects all symmetries of WTe$_2$. We rewrite the Hamiltonian here for convenience
\begin{align}
    \mc H^{\pm}_n( \bm{\delta k}) = -( \pm v_x \delta k_x \tilde\sigma_x^{\pm} + v_y \delta k_y\tilde\sigma^{\pm}_y \pm V_{\text{so}} s_z \sigma_z )  - \mu,
\end{align}
The screw symmetry $\mathcal{C}_{2x}$ maps one Fermi surface to itself, and $(\delta k_x, \delta k_y) \rightarrow (\delta k_x, -\delta k_y)$. 
For the Hamiltonian to be invariant $\tilde \sigma^\pm_\alpha$, $\alpha = x, y$ should be defined such that
\begin{align}
    &\mc C^\dagger_{2x} \tilde \sigma^{\pm}_x \mc C_{2x}  =  \tilde \sigma^{\pm}_x \nonumber \\ 
    &\mc C^\dagger_{2x} \tilde \sigma^{\pm}_y \mc C_{2x}  =  -\tilde \sigma^{\pm}_y.
\end{align}
Being a screw symmetry, the matrix representation of $\mc C_{2x}$ depends on momentum, and we rewrite it in the following form,
\begin{align}
    C_{2x} 
    = s_x(1 + e^{ik_x}) (\sigma_x + \sigma_y \tan(k_x/2)).
\end{align}
This symmetry is effectively a reflection about the axis $(1, \tan(k_x/2))$ in the $xy$-plane. 
This line makes an angle $\theta$ with the $x$-axis, such that $\tan(\theta) = \tan(k_x/2)$, or $\theta = k_x/2$. 
To maintain the symmetry, $\tilde \sigma_x $ needs to be parallel to this line, and $\tilde \sigma_y $ needs to be perpendicular to it. 

To maintain the simple form of the Dirac Hamiltonian, we approximate $\mc C_{2x}$ to be constant on each Fermi surface and evaluated at the centers $\pm K$. {It is straightforward to check that the following form for $\tilde\sigma_\alpha$, $\alpha=x,y$, preserves the screw symmetry.}

\begin{align}
    \tilde \sigma^+_{\alpha} = e^{i  K \sigma_z /2} \sigma_{\alpha} e^{-i  K \sigma_z /2}, \nonumber \\
    \tilde \sigma^-_{\alpha} = e^{-i  K \sigma_z /2} \sigma_{\alpha} e^{i  K \sigma_z /2} 
\end{align}

\section{Cancellation of first three second-order diagrams}
\label{app:canelation_of_not_cross_terms}
There are four second-order terms contributing to the scattering vertex. However, three of them cancel each other out. 
Using
\begin{align}
    \raisebox{-35pt}{
    \tikz[line join = round]{ 
    \draw[thick,->-] (-1.2, 0.7) node[anchor = south] {} -- (0,0.7)  node[anchor = south, pos=0.3] {$\bm k, \alpha_1$};
    \draw[thick, ->-] (-1.2, -0.7) node[anchor = north] {} -- (0,-0.7) node[anchor = north, pos=0.3] {$-\bm k, \alpha_2$}; 
    \draw[thick, decorate, decoration={snake, segment length=2.43mm}] (0,0.7) -- (0,-0.7cm);
    \draw[thick,->-] (0, 0.7) -- (1.2,0.7) node[anchor = south, pos=0.7] {$\bm k', \alpha'_1$}  ;
    \draw[thick, ->-] (0,-0.7) -- (1.2,-0.7)  node[anchor = north, pos=0.6] {$-\bm k', \alpha'_2$};  
    \node at (1.9,0) [scale=1] {$=$};
    }}
    &&  U_0 \delta_{\alpha_1 \alpha'_1} \delta_{\alpha_2 \alpha'_2}  \delta_{\alpha_1 \alpha_2}
\end{align}
we can evaluate the following diagrams as 
\begin{widetext}
    \begin{align}
        \raisebox{-35pt}{
        \tikz[line join = round]{ 
        \draw[thick,->-] (-1., 0.7) -- (0,0.7) node[anchor = south, pos=0.3] {$\bm k, \alpha_1$};
        \draw[thick,->-] (0,0.7) to [out=-90, in=180] (0.6, 0);
        \draw[thick,-->-] (0.6, 0) to [out=0, in=-90] (1.2, 0.7)  node[anchor=south] {};
        \draw[thick, ->-] (-1., -0.7) -- (0.0,-0.7) node[anchor = north, pos=0.3] {$-\bm k, \alpha_2$};
        \draw[thick,] (0,-0.7) -- (1.2, -0.7) node[anchor=north, pos=0.5] {}; 
        \draw[thick, decorate, decoration={snake, segment length=2.43mm}] (0,0.7) -- (1.2, 0.7);
        \draw[thick, decorate, decoration={snake, segment length=2.43mm}] (0.6,-0.7) -- (0.6, 0);
        \draw[thick,->-] (1.2, 0.7) -- (2,0.7) node[anchor = south, pos=0.5] {$\bm k', \alpha'_1$}   ;
        \draw[thick, ->-] (1.2,-0.7) -- (2,-0.7) node[anchor = north, pos=0.5] {$-\bm k', \alpha'_2$}; 
        \node at (2.6,0) [scale=1] {$=$};
        \begin{scope}[xshift=4.1cm]
            \draw[thick,->-] (-1., -0.7) -- (0,-0.7) node[anchor = north, pos=0.3] {$-\bm k, \alpha_2$}  ;
            \draw[thick,->-] (0,-0.7) to [out=90, in=180] (0.6, 0)  node[anchor=north] {};
            \draw[thick,-->-] (0.6, 0) to [out=0, in=90] (1.2, -0.7) ;
            \draw[thick, ->-] (-1., 0.7) -- (0.0,0.7) node[anchor = south, pos=0.3] {$\bm k, \alpha_1$};
            \draw[thick,] (0,0.7) -- (1.2, 0.7) node[anchor=south, pos=0.5] {}; 
            \draw[thick, decorate, decoration={snake, segment length=2.43mm}] (0,-0.7) -- (1.2,-0.7);
            \draw[thick, decorate, decoration={snake, segment length=2.43mm}] (0.6,0.7) -- (0.6, 0);
            \draw[thick,->-] (1.2, -0.7) -- (2,-0.7) node[anchor = north, pos=0.5] {$-\bm k', \alpha'_2$} ;
            \draw[thick, ->-] (1.2,0.7) -- (2,0.7)  node[anchor = south, pos=0.5] {$\bm k', \alpha'_1$}; 
            \node at (2.6,0) [scale=1] {$=$};
        \end{scope}
        }}
        && U_0^2 \delta_{\alpha_1 \alpha'_1} \delta_{\alpha_2 \alpha'_2}  \int \frac{d p}{(2\pi)^3} G^{\alpha_2 \alpha_1}(p) G^{\alpha_1 \alpha_2}(p-q), \\ 
        \raisebox{-35pt}{
            \tikz[line join = round]{ 
                \draw[thick,->-] (-1.2, -0.7) -- (-0.4,-0.7)node[anchor = north, pos=0.3] {$-\bm k, \alpha_2$} ;
                \draw[thick] (-0.4, -0.7) -- (1.2,-0.7);
                \draw[thick,->-] (1.2, -0.7) -- (2,-0.7) node[anchor = north, pos=0.5] {$-\bm k', \alpha'_2$} ;
                \draw[thick, decorate, decoration={snake, segment length=2.43mm}] (-0.4,0.7)  -- (-0.4, 0);
                \draw[thick, decorate, decoration={snake, segment length=2.43mm}] (1.2,0)-- (1.2, -0.7) node[anchor=north] {};
                \draw[thick,->-] (-0.4, 0) to [out=45, in=145] (1.2,0)  node[anchor=west] {};
                \draw[thick,->-] (1.2, 0) to [out=225, in=-45] (-0.4,0)  node[anchor= east] {};
                \draw[thick,->-] (-1.2, 0.7) -- (-0.4,0.7) node[anchor = south, pos=0.3] {$\bm k, \alpha_1$};
                \draw[thick] (-0.4, 0.7) -- (1.2,0.7);
                \draw[thick,->-] (1.2, 0.7) -- (2,0.7)  node[anchor = south, pos=0.5] {$\bm k', \alpha'_1$} ;
                \node at (4.5,0) [scale=1] {$= $};
            }}
            && -2 U_0^2 \delta_{\alpha_1 \alpha'_1} \delta_{\alpha_2 \alpha'_2}  \int \frac{d p}{(2\pi)^3} G^{\alpha_2 \alpha_1}(p) G^{\alpha_1 \alpha_2}(p-q).
    \end{align}
\end{widetext}
Here $q = k' - k$ and the $-2$ factor for the bubble diagram is due to the fermionic loop, and the spin degeneracy.  
We thus see that these three diagrams cancel each other out.

\bibliographystyle{apsrev4-2}
\bibliography{references}

\begin{thebibliography}{68}%
\makeatletter
\providecommand \@ifxundefined [1]{%
 \@ifx{#1\undefined}
}%
\providecommand \@ifnum [1]{%
 \ifnum #1\expandafter \@firstoftwo
 \else \expandafter \@secondoftwo
 \fi
}%
\providecommand \@ifx [1]{%
 \ifx #1\expandafter \@firstoftwo
 \else \expandafter \@secondoftwo
 \fi
}%
\providecommand \natexlab [1]{#1}%
\providecommand \enquote  [1]{``#1''}%
\providecommand \bibnamefont  [1]{#1}%
\providecommand \bibfnamefont [1]{#1}%
\providecommand \citenamefont [1]{#1}%
\providecommand \href@noop [0]{\@secondoftwo}%
\providecommand \href [0]{\begingroup \@sanitize@url \@href}%
\providecommand \@href[1]{\@@startlink{#1}\@@href}%
\providecommand \@@href[1]{\endgroup#1\@@endlink}%
\providecommand \@sanitize@url [0]{\catcode `\\12\catcode `\$12\catcode
  `\&12\catcode `\#12\catcode `\^12\catcode `\_12\catcode `\%12\relax}%
\providecommand \@@startlink[1]{}%
\providecommand \@@endlink[0]{}%
\providecommand \url  [0]{\begingroup\@sanitize@url \@url }%
\providecommand \@url [1]{\endgroup\@href {#1}{\urlprefix }}%
\providecommand \urlprefix  [0]{URL }%
\providecommand \Eprint [0]{\href }%
\providecommand \doibase [0]{https://doi.org/}%
\providecommand \selectlanguage [0]{\@gobble}%
\providecommand \bibinfo  [0]{\@secondoftwo}%
\providecommand \bibfield  [0]{\@secondoftwo}%
\providecommand \translation [1]{[#1]}%
\providecommand \BibitemOpen [0]{}%
\providecommand \bibitemStop [0]{}%
\providecommand \bibitemNoStop [0]{.\EOS\space}%
\providecommand \EOS [0]{\spacefactor3000\relax}%
\providecommand \BibitemShut  [1]{\csname bibitem#1\endcsname}%
\let\auto@bib@innerbib\@empty
\bibitem [{\citenamefont {Hasan}\ and\ \citenamefont
  {Kane}(2010)}]{Hasan_Kane_2010}%
  \BibitemOpen
  \bibfield  {author} {\bibinfo {author} {\bibfnamefont {M.~Z.}\ \bibnamefont
  {Hasan}}\ and\ \bibinfo {author} {\bibfnamefont {C.~L.}\ \bibnamefont
  {Kane}},\ }\href {https://doi.org/10.1103/RevModPhys.82.3045} {\bibfield
  {journal} {\bibinfo  {journal} {Reviews of Modern Physics}\ }\textbf
  {\bibinfo {volume} {82}},\ \bibinfo {pages} {3045–3067} (\bibinfo {year}
  {2010})},\ \bibinfo {note} {arXiv: 1002.3895}\BibitemShut {NoStop}%
\bibitem [{\citenamefont {Qi}\ and\ \citenamefont
  {Zhang}(2011)}]{Qi_Zhang_2011}%
  \BibitemOpen
  \bibfield  {author} {\bibinfo {author} {\bibfnamefont {X.-L.}\ \bibnamefont
  {Qi}}\ and\ \bibinfo {author} {\bibfnamefont {S.-C.}\ \bibnamefont {Zhang}},\
  }\href {https://doi.org/10.1103/RevModPhys.83.1057} {\bibfield  {journal}
  {\bibinfo  {journal} {Reviews of Modern Physics}\ }\textbf {\bibinfo {volume}
  {83}},\ \bibinfo {pages} {1057–1110} (\bibinfo {year} {2011})}\BibitemShut
  {NoStop}%
\bibitem [{\citenamefont {Chiu}\ \emph {et~al.}(2016)\citenamefont {Chiu},
  \citenamefont {Teo}, \citenamefont {Schnyder},\ and\ \citenamefont
  {Ryu}}]{Chiu_2016}%
  \BibitemOpen
  \bibfield  {author} {\bibinfo {author} {\bibfnamefont {C.-K.}\ \bibnamefont
  {Chiu}}, \bibinfo {author} {\bibfnamefont {J.~C.}\ \bibnamefont {Teo}},
  \bibinfo {author} {\bibfnamefont {A.~P.}\ \bibnamefont {Schnyder}},\ and\
  \bibinfo {author} {\bibfnamefont {S.}~\bibnamefont {Ryu}},\ }\bibfield
  {journal} {\bibinfo  {journal} {Reviews of Modern Physics}\ }\textbf
  {\bibinfo {volume} {88}},\ \href
  {https://doi.org/10.1103/revmodphys.88.035005} {10.1103/revmodphys.88.035005}
  (\bibinfo {year} {2016})\BibitemShut {NoStop}%
\bibitem [{Wit(2016)}]{Witten_2016}%
  \BibitemOpen
  \href {https://doi.org/10.1393/ncr/i2016-10125-3} {\bibfield  {journal}
  {\bibinfo  {journal} {La Rivista del Nuovo Cimento}\ }\textbf {\bibinfo
  {volume} {39}},\ \bibinfo {pages} {313–370} (\bibinfo {year}
  {2016})}\BibitemShut {NoStop}%
\bibitem [{\citenamefont {Bernevig}\ \emph {et~al.}(2006)\citenamefont
  {Bernevig}, \citenamefont {Hughes},\ and\ \citenamefont
  {Zhang}}]{Bernevig_2006}%
  \BibitemOpen
  \bibfield  {author} {\bibinfo {author} {\bibfnamefont {B.~A.}\ \bibnamefont
  {Bernevig}}, \bibinfo {author} {\bibfnamefont {T.~L.}\ \bibnamefont
  {Hughes}},\ and\ \bibinfo {author} {\bibfnamefont {S.-C.}\ \bibnamefont
  {Zhang}},\ }\href {https://doi.org/10.1126/science.1133734} {\bibfield
  {journal} {\bibinfo  {journal} {Science}\ }\textbf {\bibinfo {volume}
  {314}},\ \bibinfo {pages} {1757} (\bibinfo {year} {2006})}\BibitemShut
  {NoStop}%
\bibitem [{\citenamefont {König}\ \emph {et~al.}(2007)\citenamefont {König},
  \citenamefont {Wiedmann}, \citenamefont {Brüne}, \citenamefont {Roth},
  \citenamefont {Buhmann}, \citenamefont {Molenkamp}, \citenamefont {Qi},\ and\
  \citenamefont {Zhang}}]{Markus_2007}%
  \BibitemOpen
  \bibfield  {author} {\bibinfo {author} {\bibfnamefont {M.}~\bibnamefont
  {König}}, \bibinfo {author} {\bibfnamefont {S.}~\bibnamefont {Wiedmann}},
  \bibinfo {author} {\bibfnamefont {C.}~\bibnamefont {Brüne}}, \bibinfo
  {author} {\bibfnamefont {A.}~\bibnamefont {Roth}}, \bibinfo {author}
  {\bibfnamefont {H.}~\bibnamefont {Buhmann}}, \bibinfo {author} {\bibfnamefont
  {L.~W.}\ \bibnamefont {Molenkamp}}, \bibinfo {author} {\bibfnamefont {X.-L.}\
  \bibnamefont {Qi}},\ and\ \bibinfo {author} {\bibfnamefont {S.-C.}\
  \bibnamefont {Zhang}},\ }\href {https://doi.org/10.1126/science.1148047}
  {\bibfield  {journal} {\bibinfo  {journal} {Science}\ }\textbf {\bibinfo
  {volume} {318}},\ \bibinfo {pages} {766} (\bibinfo {year} {2007})},\ \Eprint
  {https://arxiv.org/abs/https://www.science.org/doi/pdf/10.1126/science.1148047}
  {https://www.science.org/doi/pdf/10.1126/science.1148047} \BibitemShut
  {NoStop}%
\bibitem [{\citenamefont {Nowack}\ \emph {et~al.}(2013)\citenamefont {Nowack},
  \citenamefont {Spanton}, \citenamefont {Baenninger}, \citenamefont {König},
  \citenamefont {Kirtley}, \citenamefont {Kalisky}, \citenamefont {Ames},
  \citenamefont {Leubner}, \citenamefont {Brüne}, \citenamefont {Buhmann},
  \citenamefont {Molenkamp}, \citenamefont {Goldhaber-Gordon},\ and\
  \citenamefont {Moler}}]{Nowack_2013}%
  \BibitemOpen
  \bibfield  {author} {\bibinfo {author} {\bibfnamefont {K.~C.}\ \bibnamefont
  {Nowack}}, \bibinfo {author} {\bibfnamefont {E.~M.}\ \bibnamefont {Spanton}},
  \bibinfo {author} {\bibfnamefont {M.}~\bibnamefont {Baenninger}}, \bibinfo
  {author} {\bibfnamefont {M.}~\bibnamefont {König}}, \bibinfo {author}
  {\bibfnamefont {J.~R.}\ \bibnamefont {Kirtley}}, \bibinfo {author}
  {\bibfnamefont {B.}~\bibnamefont {Kalisky}}, \bibinfo {author} {\bibfnamefont
  {C.}~\bibnamefont {Ames}}, \bibinfo {author} {\bibfnamefont {P.}~\bibnamefont
  {Leubner}}, \bibinfo {author} {\bibfnamefont {C.}~\bibnamefont {Brüne}},
  \bibinfo {author} {\bibfnamefont {H.}~\bibnamefont {Buhmann}}, \bibinfo
  {author} {\bibfnamefont {L.~W.}\ \bibnamefont {Molenkamp}}, \bibinfo {author}
  {\bibfnamefont {D.}~\bibnamefont {Goldhaber-Gordon}},\ and\ \bibinfo {author}
  {\bibfnamefont {K.~A.}\ \bibnamefont {Moler}},\ }\href
  {https://doi.org/10.1038/nmat3682} {\bibfield  {journal} {\bibinfo  {journal}
  {Nature Materials}\ }\textbf {\bibinfo {volume} {12}},\ \bibinfo {pages}
  {787} (\bibinfo {year} {2013})}\BibitemShut {NoStop}%
\bibitem [{\citenamefont {Deacon}\ \emph {et~al.}(2017)\citenamefont {Deacon},
  \citenamefont {Wiedenmann}, \citenamefont {Bocquillon}, \citenamefont
  {Dom\'{\i}nguez}, \citenamefont {Klapwijk}, \citenamefont {Leubner},
  \citenamefont {Br\"une}, \citenamefont {Hankiewicz}, \citenamefont {Tarucha},
  \citenamefont {Ishibashi}, \citenamefont {Buhmann},\ and\ \citenamefont
  {Molenkamp}}]{Deacon_2017}%
  \BibitemOpen
  \bibfield  {author} {\bibinfo {author} {\bibfnamefont {R.~S.}\ \bibnamefont
  {Deacon}}, \bibinfo {author} {\bibfnamefont {J.}~\bibnamefont {Wiedenmann}},
  \bibinfo {author} {\bibfnamefont {E.}~\bibnamefont {Bocquillon}}, \bibinfo
  {author} {\bibfnamefont {F.}~\bibnamefont {Dom\'{\i}nguez}}, \bibinfo
  {author} {\bibfnamefont {T.~M.}\ \bibnamefont {Klapwijk}}, \bibinfo {author}
  {\bibfnamefont {P.}~\bibnamefont {Leubner}}, \bibinfo {author} {\bibfnamefont
  {C.}~\bibnamefont {Br\"une}}, \bibinfo {author} {\bibfnamefont {E.~M.}\
  \bibnamefont {Hankiewicz}}, \bibinfo {author} {\bibfnamefont
  {S.}~\bibnamefont {Tarucha}}, \bibinfo {author} {\bibfnamefont
  {K.}~\bibnamefont {Ishibashi}}, \bibinfo {author} {\bibfnamefont
  {H.}~\bibnamefont {Buhmann}},\ and\ \bibinfo {author} {\bibfnamefont {L.~W.}\
  \bibnamefont {Molenkamp}},\ }\href
  {https://doi.org/10.1103/PhysRevX.7.021011} {\bibfield  {journal} {\bibinfo
  {journal} {Phys. Rev. X}\ }\textbf {\bibinfo {volume} {7}},\ \bibinfo {pages}
  {021011} (\bibinfo {year} {2017})}\BibitemShut {NoStop}%
\bibitem [{\citenamefont {Shi}\ \emph {et~al.}(2019)\citenamefont {Shi},
  \citenamefont {Kahn}, \citenamefont {Niu}, \citenamefont {Fei}, \citenamefont
  {Sun}, \citenamefont {Cai}, \citenamefont {Francisco}, \citenamefont {Wu},
  \citenamefont {Shen}, \citenamefont {Xu}, \citenamefont {Cobden},\ and\
  \citenamefont {Cui}}]{Yanmeng_2019}%
  \BibitemOpen
  \bibfield  {author} {\bibinfo {author} {\bibfnamefont {Y.}~\bibnamefont
  {Shi}}, \bibinfo {author} {\bibfnamefont {J.}~\bibnamefont {Kahn}}, \bibinfo
  {author} {\bibfnamefont {B.}~\bibnamefont {Niu}}, \bibinfo {author}
  {\bibfnamefont {Z.}~\bibnamefont {Fei}}, \bibinfo {author} {\bibfnamefont
  {B.}~\bibnamefont {Sun}}, \bibinfo {author} {\bibfnamefont {X.}~\bibnamefont
  {Cai}}, \bibinfo {author} {\bibfnamefont {B.~A.}\ \bibnamefont {Francisco}},
  \bibinfo {author} {\bibfnamefont {D.}~\bibnamefont {Wu}}, \bibinfo {author}
  {\bibfnamefont {Z.-X.}\ \bibnamefont {Shen}}, \bibinfo {author}
  {\bibfnamefont {X.}~\bibnamefont {Xu}}, \bibinfo {author} {\bibfnamefont
  {D.~H.}\ \bibnamefont {Cobden}},\ and\ \bibinfo {author} {\bibfnamefont
  {Y.-T.}\ \bibnamefont {Cui}},\ }\href
  {https://doi.org/10.1126/sciadv.aat8799} {\bibfield  {journal} {\bibinfo
  {journal} {Science Advances}\ }\textbf {\bibinfo {volume} {5}},\ \bibinfo
  {pages} {eaat8799} (\bibinfo {year} {2019})},\ \Eprint
  {https://arxiv.org/abs/https://www.science.org/doi/pdf/10.1126/sciadv.aat8799}
  {https://www.science.org/doi/pdf/10.1126/sciadv.aat8799} \BibitemShut
  {NoStop}%
\bibitem [{\citenamefont {Wu}\ \emph {et~al.}(2018)\citenamefont {Wu},
  \citenamefont {Fatemi}, \citenamefont {Gibson}, \citenamefont {Watanabe},
  \citenamefont {Taniguchi}, \citenamefont {Cava},\ and\ \citenamefont
  {Jarillo-Herrero}}]{Sanfeng_2018}%
  \BibitemOpen
  \bibfield  {author} {\bibinfo {author} {\bibfnamefont {S.}~\bibnamefont
  {Wu}}, \bibinfo {author} {\bibfnamefont {V.}~\bibnamefont {Fatemi}}, \bibinfo
  {author} {\bibfnamefont {Q.~D.}\ \bibnamefont {Gibson}}, \bibinfo {author}
  {\bibfnamefont {K.}~\bibnamefont {Watanabe}}, \bibinfo {author}
  {\bibfnamefont {T.}~\bibnamefont {Taniguchi}}, \bibinfo {author}
  {\bibfnamefont {R.~J.}\ \bibnamefont {Cava}},\ and\ \bibinfo {author}
  {\bibfnamefont {P.}~\bibnamefont {Jarillo-Herrero}},\ }\href
  {https://doi.org/10.1126/science.aan6003} {\bibfield  {journal} {\bibinfo
  {journal} {Science}\ }\textbf {\bibinfo {volume} {359}},\ \bibinfo {pages}
  {76} (\bibinfo {year} {2018})},\ \Eprint
  {https://arxiv.org/abs/https://www.science.org/doi/pdf/10.1126/science.aan6003}
  {https://www.science.org/doi/pdf/10.1126/science.aan6003} \BibitemShut
  {NoStop}%
\bibitem [{\citenamefont {Tang}\ \emph {et~al.}(2017)\citenamefont {Tang},
  \citenamefont {Zhang}, \citenamefont {Wong}, \citenamefont {Pedramrazi},
  \citenamefont {Tsai}, \citenamefont {Jia}, \citenamefont {Moritz},
  \citenamefont {Claassen}, \citenamefont {Ryu}, \citenamefont {Kahn},
  \citenamefont {Jiang}, \citenamefont {Yan}, \citenamefont {Hashimoto},
  \citenamefont {Lu}, \citenamefont {Moore}, \citenamefont {Hwang},
  \citenamefont {Hwang}, \citenamefont {Hussain}, \citenamefont {Chen},
  \citenamefont {Ugeda}, \citenamefont {Liu}, \citenamefont {Xie},
  \citenamefont {Devereaux}, \citenamefont {Crommie}, \citenamefont {Mo},\ and\
  \citenamefont {Shen}}]{Tang_2017}%
  \BibitemOpen
  \bibfield  {author} {\bibinfo {author} {\bibfnamefont {S.}~\bibnamefont
  {Tang}}, \bibinfo {author} {\bibfnamefont {C.}~\bibnamefont {Zhang}},
  \bibinfo {author} {\bibfnamefont {D.}~\bibnamefont {Wong}}, \bibinfo {author}
  {\bibfnamefont {Z.}~\bibnamefont {Pedramrazi}}, \bibinfo {author}
  {\bibfnamefont {H.-Z.}\ \bibnamefont {Tsai}}, \bibinfo {author}
  {\bibfnamefont {C.}~\bibnamefont {Jia}}, \bibinfo {author} {\bibfnamefont
  {B.}~\bibnamefont {Moritz}}, \bibinfo {author} {\bibfnamefont
  {M.}~\bibnamefont {Claassen}}, \bibinfo {author} {\bibfnamefont
  {H.}~\bibnamefont {Ryu}}, \bibinfo {author} {\bibfnamefont {S.}~\bibnamefont
  {Kahn}}, \bibinfo {author} {\bibfnamefont {J.}~\bibnamefont {Jiang}},
  \bibinfo {author} {\bibfnamefont {H.}~\bibnamefont {Yan}}, \bibinfo {author}
  {\bibfnamefont {M.}~\bibnamefont {Hashimoto}}, \bibinfo {author}
  {\bibfnamefont {D.}~\bibnamefont {Lu}}, \bibinfo {author} {\bibfnamefont
  {R.~G.}\ \bibnamefont {Moore}}, \bibinfo {author} {\bibfnamefont {C.-C.}\
  \bibnamefont {Hwang}}, \bibinfo {author} {\bibfnamefont {C.}~\bibnamefont
  {Hwang}}, \bibinfo {author} {\bibfnamefont {Z.}~\bibnamefont {Hussain}},
  \bibinfo {author} {\bibfnamefont {Y.}~\bibnamefont {Chen}}, \bibinfo {author}
  {\bibfnamefont {M.~M.}\ \bibnamefont {Ugeda}}, \bibinfo {author}
  {\bibfnamefont {Z.}~\bibnamefont {Liu}}, \bibinfo {author} {\bibfnamefont
  {X.}~\bibnamefont {Xie}}, \bibinfo {author} {\bibfnamefont {T.~P.}\
  \bibnamefont {Devereaux}}, \bibinfo {author} {\bibfnamefont {M.~F.}\
  \bibnamefont {Crommie}}, \bibinfo {author} {\bibfnamefont {S.-K.}\
  \bibnamefont {Mo}},\ and\ \bibinfo {author} {\bibfnamefont {Z.-X.}\
  \bibnamefont {Shen}},\ }\href {https://doi.org/10.1038/nphys4174} {\bibfield
  {journal} {\bibinfo  {journal} {Nature Physics}\ }\textbf {\bibinfo {volume}
  {13}},\ \bibinfo {pages} {683–687} (\bibinfo {year} {2017})}\BibitemShut
  {NoStop}%
\bibitem [{\citenamefont {Fei}\ \emph {et~al.}(2017)\citenamefont {Fei},
  \citenamefont {Palomaki}, \citenamefont {Wu}, \citenamefont {Zhao},
  \citenamefont {Cai}, \citenamefont {Sun}, \citenamefont {Nguyen},
  \citenamefont {Finney}, \citenamefont {Xu},\ and\ \citenamefont
  {Cobden}}]{Fei_2017}%
  \BibitemOpen
  \bibfield  {author} {\bibinfo {author} {\bibfnamefont {Z.}~\bibnamefont
  {Fei}}, \bibinfo {author} {\bibfnamefont {T.}~\bibnamefont {Palomaki}},
  \bibinfo {author} {\bibfnamefont {S.}~\bibnamefont {Wu}}, \bibinfo {author}
  {\bibfnamefont {W.}~\bibnamefont {Zhao}}, \bibinfo {author} {\bibfnamefont
  {X.}~\bibnamefont {Cai}}, \bibinfo {author} {\bibfnamefont {B.}~\bibnamefont
  {Sun}}, \bibinfo {author} {\bibfnamefont {P.}~\bibnamefont {Nguyen}},
  \bibinfo {author} {\bibfnamefont {J.}~\bibnamefont {Finney}}, \bibinfo
  {author} {\bibfnamefont {X.}~\bibnamefont {Xu}},\ and\ \bibinfo {author}
  {\bibfnamefont {D.~H.}\ \bibnamefont {Cobden}},\ }\href
  {https://doi.org/10.1038/nphys4091} {\bibfield  {journal} {\bibinfo
  {journal} {Nature Physics}\ }\textbf {\bibinfo {volume} {13}},\ \bibinfo
  {pages} {677–682} (\bibinfo {year} {2017})}\BibitemShut {NoStop}%
\bibitem [{\citenamefont {Jia}\ \emph {et~al.}(2017)\citenamefont {Jia},
  \citenamefont {Song}, \citenamefont {Li}, \citenamefont {Ran}, \citenamefont
  {Lu}, \citenamefont {Zheng}, \citenamefont {Zhu}, \citenamefont {Shi},
  \citenamefont {Sun}, \citenamefont {Wen}, \citenamefont {Xing},\ and\
  \citenamefont {Li}}]{Jia_2017}%
  \BibitemOpen
  \bibfield  {author} {\bibinfo {author} {\bibfnamefont {Z.-Y.}\ \bibnamefont
  {Jia}}, \bibinfo {author} {\bibfnamefont {Y.-H.}\ \bibnamefont {Song}},
  \bibinfo {author} {\bibfnamefont {X.-B.}\ \bibnamefont {Li}}, \bibinfo
  {author} {\bibfnamefont {K.}~\bibnamefont {Ran}}, \bibinfo {author}
  {\bibfnamefont {P.}~\bibnamefont {Lu}}, \bibinfo {author} {\bibfnamefont
  {H.-J.}\ \bibnamefont {Zheng}}, \bibinfo {author} {\bibfnamefont {X.-Y.}\
  \bibnamefont {Zhu}}, \bibinfo {author} {\bibfnamefont {Z.-Q.}\ \bibnamefont
  {Shi}}, \bibinfo {author} {\bibfnamefont {J.}~\bibnamefont {Sun}}, \bibinfo
  {author} {\bibfnamefont {J.}~\bibnamefont {Wen}}, \bibinfo {author}
  {\bibfnamefont {D.}~\bibnamefont {Xing}},\ and\ \bibinfo {author}
  {\bibfnamefont {S.-C.}\ \bibnamefont {Li}},\ }\href
  {https://doi.org/10.1103/PhysRevB.96.041108} {\bibfield  {journal} {\bibinfo
  {journal} {Phys. Rev. B}\ }\textbf {\bibinfo {volume} {96}},\ \bibinfo
  {pages} {041108} (\bibinfo {year} {2017})}\BibitemShut {NoStop}%
\bibitem [{\citenamefont {Peng}\ \emph {et~al.}(2017)\citenamefont {Peng},
  \citenamefont {Yuan}, \citenamefont {Li}, \citenamefont {Yang}, \citenamefont
  {Xian}, \citenamefont {Yi}, \citenamefont {Shi},\ and\ \citenamefont
  {Fu}}]{Peng_2017}%
  \BibitemOpen
  \bibfield  {author} {\bibinfo {author} {\bibfnamefont {L.}~\bibnamefont
  {Peng}}, \bibinfo {author} {\bibfnamefont {Y.}~\bibnamefont {Yuan}}, \bibinfo
  {author} {\bibfnamefont {G.}~\bibnamefont {Li}}, \bibinfo {author}
  {\bibfnamefont {X.}~\bibnamefont {Yang}}, \bibinfo {author} {\bibfnamefont
  {J.-J.}\ \bibnamefont {Xian}}, \bibinfo {author} {\bibfnamefont {C.-J.}\
  \bibnamefont {Yi}}, \bibinfo {author} {\bibfnamefont {Y.-G.}\ \bibnamefont
  {Shi}},\ and\ \bibinfo {author} {\bibfnamefont {Y.-S.}\ \bibnamefont {Fu}},\
  }\href {https://doi.org/10.1038/s41467-017-00745-8} {\bibfield  {journal}
  {\bibinfo  {journal} {Nature Communications}\ }\textbf {\bibinfo {volume}
  {8}},\ \bibinfo {pages} {659} (\bibinfo {year} {2017})}\BibitemShut {NoStop}%
\bibitem [{\citenamefont {Lau}\ \emph {et~al.}(2019)\citenamefont {Lau},
  \citenamefont {Ray}, \citenamefont {Varjas},\ and\ \citenamefont
  {Akhmerov}}]{Akhmerov_2019}%
  \BibitemOpen
  \bibfield  {author} {\bibinfo {author} {\bibfnamefont {A.}~\bibnamefont
  {Lau}}, \bibinfo {author} {\bibfnamefont {R.}~\bibnamefont {Ray}}, \bibinfo
  {author} {\bibfnamefont {D.}~\bibnamefont {Varjas}},\ and\ \bibinfo {author}
  {\bibfnamefont {A.~R.}\ \bibnamefont {Akhmerov}},\ }\href
  {https://doi.org/10.1103/PhysRevMaterials.3.054206} {\bibfield  {journal}
  {\bibinfo  {journal} {Phys. Rev. Mater.}\ }\textbf {\bibinfo {volume} {3}},\
  \bibinfo {pages} {054206} (\bibinfo {year} {2019})}\BibitemShut {NoStop}%
\bibitem [{\citenamefont {Sarma}\ \emph {et~al.}(2015)\citenamefont {Sarma},
  \citenamefont {Freedman},\ and\ \citenamefont {Nayak}}]{Sarma_2015}%
  \BibitemOpen
  \bibfield  {author} {\bibinfo {author} {\bibfnamefont {S.~D.}\ \bibnamefont
  {Sarma}}, \bibinfo {author} {\bibfnamefont {M.}~\bibnamefont {Freedman}},\
  and\ \bibinfo {author} {\bibfnamefont {C.}~\bibnamefont {Nayak}},\ }\bibfield
   {journal} {\bibinfo  {journal} {npj Quantum Information}\ }\textbf {\bibinfo
  {volume} {1}},\ \href {https://doi.org/10.1038/npjqi.2015.1}
  {10.1038/npjqi.2015.1} (\bibinfo {year} {2015})\BibitemShut {NoStop}%
\bibitem [{\citenamefont {Beenakker}(2013)}]{Beenakker_2013}%
  \BibitemOpen
  \bibfield  {author} {\bibinfo {author} {\bibfnamefont {C.}~\bibnamefont
  {Beenakker}},\ }\href
  {https://doi.org/10.1146/annurev-conmatphys-030212-184337} {\bibfield
  {journal} {\bibinfo  {journal} {Annual Review of Condensed Matter Physics}\
  }\textbf {\bibinfo {volume} {4}},\ \bibinfo {pages} {113} (\bibinfo {year}
  {2013})}\BibitemShut {NoStop}%
\bibitem [{\citenamefont {Yan}(2019)}]{Yan_2019}%
  \BibitemOpen
  \bibfield  {author} {\bibinfo {author} {\bibfnamefont {Z.}~\bibnamefont
  {Yan}},\ }\href {https://doi.org/10.1103/PhysRevLett.123.177001} {\bibfield
  {journal} {\bibinfo  {journal} {Physical Review Letters}\ }\textbf {\bibinfo
  {volume} {123}},\ \bibinfo {pages} {177001} (\bibinfo {year}
  {2019})}\BibitemShut {NoStop}%
\bibitem [{\citenamefont {Benalcazar}\ \emph {et~al.}(2017)\citenamefont
  {Benalcazar}, \citenamefont {Bernevig},\ and\ \citenamefont
  {Hughes}}]{Benalcazar_2017}%
  \BibitemOpen
  \bibfield  {author} {\bibinfo {author} {\bibfnamefont {W.~A.}\ \bibnamefont
  {Benalcazar}}, \bibinfo {author} {\bibfnamefont {B.~A.}\ \bibnamefont
  {Bernevig}},\ and\ \bibinfo {author} {\bibfnamefont {T.~L.}\ \bibnamefont
  {Hughes}},\ }\href {https://doi.org/10.1126/science.aah6442} {\bibfield
  {journal} {\bibinfo  {journal} {Science}\ }\textbf {\bibinfo {volume}
  {357}},\ \bibinfo {pages} {61–66} (\bibinfo {year} {2017})}\BibitemShut
  {NoStop}%
\bibitem [{\citenamefont {Trifunovic}\ and\ \citenamefont
  {Brouwer}(2019)}]{Trifunovic_2019}%
  \BibitemOpen
  \bibfield  {author} {\bibinfo {author} {\bibfnamefont {L.}~\bibnamefont
  {Trifunovic}}\ and\ \bibinfo {author} {\bibfnamefont {P.~W.}\ \bibnamefont
  {Brouwer}},\ }\href {https://doi.org/10.1103/PhysRevX.9.011012} {\bibfield
  {journal} {\bibinfo  {journal} {Physical Review X}\ }\textbf {\bibinfo
  {volume} {9}},\ \bibinfo {pages} {011012} (\bibinfo {year}
  {2019})}\BibitemShut {NoStop}%
\bibitem [{\citenamefont {Geier}\ \emph {et~al.}(2018)\citenamefont {Geier},
  \citenamefont {Trifunovic}, \citenamefont {Hoskam},\ and\ \citenamefont
  {Brouwer}}]{Geier_2018}%
  \BibitemOpen
  \bibfield  {author} {\bibinfo {author} {\bibfnamefont {M.}~\bibnamefont
  {Geier}}, \bibinfo {author} {\bibfnamefont {L.}~\bibnamefont {Trifunovic}},
  \bibinfo {author} {\bibfnamefont {M.}~\bibnamefont {Hoskam}},\ and\ \bibinfo
  {author} {\bibfnamefont {P.~W.}\ \bibnamefont {Brouwer}},\ }\href
  {https://doi.org/10.1103/PhysRevB.97.205135} {\bibfield  {journal} {\bibinfo
  {journal} {Physical Review B}\ }\textbf {\bibinfo {volume} {97}},\ \bibinfo
  {pages} {205135} (\bibinfo {year} {2018})}\BibitemShut {NoStop}%
\bibitem [{\citenamefont {Schindler}\ \emph {et~al.}(2018)\citenamefont
  {Schindler}, \citenamefont {Cook}, \citenamefont {Vergniory}, \citenamefont
  {Wang}, \citenamefont {Parkin}, \citenamefont {Bernevig},\ and\ \citenamefont
  {Neupert}}]{Schindler_2018}%
  \BibitemOpen
  \bibfield  {author} {\bibinfo {author} {\bibfnamefont {F.}~\bibnamefont
  {Schindler}}, \bibinfo {author} {\bibfnamefont {A.~M.}\ \bibnamefont {Cook}},
  \bibinfo {author} {\bibfnamefont {M.~G.}\ \bibnamefont {Vergniory}}, \bibinfo
  {author} {\bibfnamefont {Z.}~\bibnamefont {Wang}}, \bibinfo {author}
  {\bibfnamefont {S.~S.~P.}\ \bibnamefont {Parkin}}, \bibinfo {author}
  {\bibfnamefont {B.~A.}\ \bibnamefont {Bernevig}},\ and\ \bibinfo {author}
  {\bibfnamefont {T.}~\bibnamefont {Neupert}},\ }\href
  {https://doi.org/10.1126/sciadv.aat0346} {\bibfield  {journal} {\bibinfo
  {journal} {Science Advances}\ }\textbf {\bibinfo {volume} {4}},\ \bibinfo
  {pages} {eaat0346} (\bibinfo {year} {2018})}\BibitemShut {NoStop}%
\bibitem [{\citenamefont {Wang}\ \emph {et~al.}(2018)\citenamefont {Wang},
  \citenamefont {Lin},\ and\ \citenamefont {Hughes}}]{Wang_2018}%
  \BibitemOpen
  \bibfield  {author} {\bibinfo {author} {\bibfnamefont {Y.}~\bibnamefont
  {Wang}}, \bibinfo {author} {\bibfnamefont {M.}~\bibnamefont {Lin}},\ and\
  \bibinfo {author} {\bibfnamefont {T.~L.}\ \bibnamefont {Hughes}},\ }\href
  {https://doi.org/10.1103/PhysRevB.98.165144} {\bibfield  {journal} {\bibinfo
  {journal} {Physical Review B}\ }\textbf {\bibinfo {volume} {98}},\ \bibinfo
  {pages} {165144} (\bibinfo {year} {2018})}\BibitemShut {NoStop}%
\bibitem [{\citenamefont {Jahin}\ \emph {et~al.}(2022)\citenamefont {Jahin},
  \citenamefont {Tiwari},\ and\ \citenamefont {Wang}}]{Jahin_2022}%
  \BibitemOpen
  \bibfield  {author} {\bibinfo {author} {\bibfnamefont {A.}~\bibnamefont
  {Jahin}}, \bibinfo {author} {\bibfnamefont {A.}~\bibnamefont {Tiwari}},\ and\
  \bibinfo {author} {\bibfnamefont {Y.}~\bibnamefont {Wang}},\ }\href
  {https://doi.org/10.21468/SciPostPhys.12.2.053} {\bibfield  {journal}
  {\bibinfo  {journal} {SciPost Physics}\ }\textbf {\bibinfo {volume} {12}},\
  \bibinfo {pages} {053} (\bibinfo {year} {2022})}\BibitemShut {NoStop}%
\bibitem [{\citenamefont {Tiwari}\ \emph {et~al.}(2020)\citenamefont {Tiwari},
  \citenamefont {Jahin},\ and\ \citenamefont {Wang}}]{Tiwari_2020}%
  \BibitemOpen
  \bibfield  {author} {\bibinfo {author} {\bibfnamefont {A.}~\bibnamefont
  {Tiwari}}, \bibinfo {author} {\bibfnamefont {A.}~\bibnamefont {Jahin}},\ and\
  \bibinfo {author} {\bibfnamefont {Y.}~\bibnamefont {Wang}},\ }\href
  {https://doi.org/10.1103/PhysRevResearch.2.043300} {\bibfield  {journal}
  {\bibinfo  {journal} {Physical Review Research}\ }\textbf {\bibinfo {volume}
  {2}},\ \bibinfo {pages} {043300} (\bibinfo {year} {2020})}\BibitemShut
  {NoStop}%
\bibitem [{\citenamefont {Langbehn}\ \emph {et~al.}(2017)\citenamefont
  {Langbehn}, \citenamefont {Peng}, \citenamefont {Trifunovic}, \citenamefont
  {von Oppen},\ and\ \citenamefont {Brouwer}}]{Langbehn_2017}%
  \BibitemOpen
  \bibfield  {author} {\bibinfo {author} {\bibfnamefont {J.}~\bibnamefont
  {Langbehn}}, \bibinfo {author} {\bibfnamefont {Y.}~\bibnamefont {Peng}},
  \bibinfo {author} {\bibfnamefont {L.}~\bibnamefont {Trifunovic}}, \bibinfo
  {author} {\bibfnamefont {F.}~\bibnamefont {von Oppen}},\ and\ \bibinfo
  {author} {\bibfnamefont {P.~W.}\ \bibnamefont {Brouwer}},\ }\href
  {https://doi.org/10.1103/PhysRevLett.119.246401} {\bibfield  {journal}
  {\bibinfo  {journal} {Phys. Rev. Lett.}\ }\textbf {\bibinfo {volume} {119}},\
  \bibinfo {pages} {246401} (\bibinfo {year} {2017})}\BibitemShut {NoStop}%
\bibitem [{\citenamefont {Zhang}\ \emph {et~al.}(2019)\citenamefont {Zhang},
  \citenamefont {Cole},\ and\ \citenamefont {Das~Sarma}}]{Zhang_2019}%
  \BibitemOpen
  \bibfield  {author} {\bibinfo {author} {\bibfnamefont {R.-X.}\ \bibnamefont
  {Zhang}}, \bibinfo {author} {\bibfnamefont {W.~S.}\ \bibnamefont {Cole}},\
  and\ \bibinfo {author} {\bibfnamefont {S.}~\bibnamefont {Das~Sarma}},\ }\href
  {https://doi.org/10.1103/PhysRevLett.122.187001} {\bibfield  {journal}
  {\bibinfo  {journal} {Phys. Rev. Lett.}\ }\textbf {\bibinfo {volume} {122}},\
  \bibinfo {pages} {187001} (\bibinfo {year} {2019})}\BibitemShut {NoStop}%
\bibitem [{\citenamefont {Hsu}\ \emph {et~al.}(2020)\citenamefont {Hsu},
  \citenamefont {Cole}, \citenamefont {Zhang},\ and\ \citenamefont
  {Sau}}]{Hsu_2020}%
  \BibitemOpen
  \bibfield  {author} {\bibinfo {author} {\bibfnamefont {Y.-T.}\ \bibnamefont
  {Hsu}}, \bibinfo {author} {\bibfnamefont {W.~S.}\ \bibnamefont {Cole}},
  \bibinfo {author} {\bibfnamefont {R.-X.}\ \bibnamefont {Zhang}},\ and\
  \bibinfo {author} {\bibfnamefont {J.~D.}\ \bibnamefont {Sau}},\ }\href
  {https://doi.org/10.1103/PhysRevLett.125.097001} {\bibfield  {journal}
  {\bibinfo  {journal} {Physical Review Letters}\ }\textbf {\bibinfo {volume}
  {125}},\ \bibinfo {pages} {097001} (\bibinfo {year} {2020})}\BibitemShut
  {NoStop}%
\bibitem [{\citenamefont {Sajadi}\ \emph {et~al.}(2018)\citenamefont {Sajadi},
  \citenamefont {Palomaki}, \citenamefont {Fei}, \citenamefont {Zhao},
  \citenamefont {Bement}, \citenamefont {Olsen}, \citenamefont {Luescher},
  \citenamefont {Xu}, \citenamefont {Folk},\ and\ \citenamefont
  {Cobden}}]{Ebrahim_2018}%
  \BibitemOpen
  \bibfield  {author} {\bibinfo {author} {\bibfnamefont {E.}~\bibnamefont
  {Sajadi}}, \bibinfo {author} {\bibfnamefont {T.}~\bibnamefont {Palomaki}},
  \bibinfo {author} {\bibfnamefont {Z.}~\bibnamefont {Fei}}, \bibinfo {author}
  {\bibfnamefont {W.}~\bibnamefont {Zhao}}, \bibinfo {author} {\bibfnamefont
  {P.}~\bibnamefont {Bement}}, \bibinfo {author} {\bibfnamefont
  {C.}~\bibnamefont {Olsen}}, \bibinfo {author} {\bibfnamefont
  {S.}~\bibnamefont {Luescher}}, \bibinfo {author} {\bibfnamefont
  {X.}~\bibnamefont {Xu}}, \bibinfo {author} {\bibfnamefont {J.~A.}\
  \bibnamefont {Folk}},\ and\ \bibinfo {author} {\bibfnamefont {D.~H.}\
  \bibnamefont {Cobden}},\ }\href {https://doi.org/10.1126/science.aar4426}
  {\bibfield  {journal} {\bibinfo  {journal} {Science}\ }\textbf {\bibinfo
  {volume} {362}},\ \bibinfo {pages} {922} (\bibinfo {year} {2018})},\ \Eprint
  {https://arxiv.org/abs/https://www.science.org/doi/pdf/10.1126/science.aar4426}
  {https://www.science.org/doi/pdf/10.1126/science.aar4426} \BibitemShut
  {NoStop}%
\bibitem [{\citenamefont {Fatemi}\ \emph {et~al.}(2018)\citenamefont {Fatemi},
  \citenamefont {Wu}, \citenamefont {Cao}, \citenamefont {Bretheau},
  \citenamefont {Gibson}, \citenamefont {Watanabe}, \citenamefont {Taniguchi},
  \citenamefont {Cava},\ and\ \citenamefont {Jarillo-Herrero}}]{Valla_2018}%
  \BibitemOpen
  \bibfield  {author} {\bibinfo {author} {\bibfnamefont {V.}~\bibnamefont
  {Fatemi}}, \bibinfo {author} {\bibfnamefont {S.}~\bibnamefont {Wu}}, \bibinfo
  {author} {\bibfnamefont {Y.}~\bibnamefont {Cao}}, \bibinfo {author}
  {\bibfnamefont {L.}~\bibnamefont {Bretheau}}, \bibinfo {author}
  {\bibfnamefont {Q.~D.}\ \bibnamefont {Gibson}}, \bibinfo {author}
  {\bibfnamefont {K.}~\bibnamefont {Watanabe}}, \bibinfo {author}
  {\bibfnamefont {T.}~\bibnamefont {Taniguchi}}, \bibinfo {author}
  {\bibfnamefont {R.~J.}\ \bibnamefont {Cava}},\ and\ \bibinfo {author}
  {\bibfnamefont {P.}~\bibnamefont {Jarillo-Herrero}},\ }\href
  {https://doi.org/10.1126/science.aar4642} {\bibfield  {journal} {\bibinfo
  {journal} {Science}\ }\textbf {\bibinfo {volume} {362}},\ \bibinfo {pages}
  {926} (\bibinfo {year} {2018})},\ \Eprint
  {https://arxiv.org/abs/https://www.science.org/doi/pdf/10.1126/science.aar4642}
  {https://www.science.org/doi/pdf/10.1126/science.aar4642} \BibitemShut
  {NoStop}%
\bibitem [{\citenamefont {Crépel}\ and\ \citenamefont {Fu}(2022)}]{Fu_2022}%
  \BibitemOpen
  \bibfield  {author} {\bibinfo {author} {\bibfnamefont {V.}~\bibnamefont
  {Crépel}}\ and\ \bibinfo {author} {\bibfnamefont {L.}~\bibnamefont {Fu}},\
  }\href {https://doi.org/10.1073/pnas.2117735119} {\bibfield  {journal}
  {\bibinfo  {journal} {Proceedings of the National Academy of Sciences}\
  }\textbf {\bibinfo {volume} {119}},\ \bibinfo {pages} {e2117735119} (\bibinfo
  {year} {2022})}\BibitemShut {NoStop}%
\bibitem [{\citenamefont {Cr\'epel}\ \emph {et~al.}(2022)\citenamefont
  {Cr\'epel}, \citenamefont {Cea}, \citenamefont {Fu},\ and\ \citenamefont
  {Guinea}}]{Francisco_2022}%
  \BibitemOpen
  \bibfield  {author} {\bibinfo {author} {\bibfnamefont {V.}~\bibnamefont
  {Cr\'epel}}, \bibinfo {author} {\bibfnamefont {T.}~\bibnamefont {Cea}},
  \bibinfo {author} {\bibfnamefont {L.}~\bibnamefont {Fu}},\ and\ \bibinfo
  {author} {\bibfnamefont {F.}~\bibnamefont {Guinea}},\ }\href
  {https://doi.org/10.1103/PhysRevB.105.094506} {\bibfield  {journal} {\bibinfo
   {journal} {Phys. Rev. B}\ }\textbf {\bibinfo {volume} {105}},\ \bibinfo
  {pages} {094506} (\bibinfo {year} {2022})}\BibitemShut {NoStop}%
\bibitem [{\citenamefont {Kohn}\ and\ \citenamefont
  {Luttinger}(1965)}]{KL_1965}%
  \BibitemOpen
  \bibfield  {author} {\bibinfo {author} {\bibfnamefont {W.}~\bibnamefont
  {Kohn}}\ and\ \bibinfo {author} {\bibfnamefont {J.~M.}\ \bibnamefont
  {Luttinger}},\ }\href {https://doi.org/10.1103/PhysRevLett.15.524} {\bibfield
   {journal} {\bibinfo  {journal} {Phys. Rev. Lett.}\ }\textbf {\bibinfo
  {volume} {15}},\ \bibinfo {pages} {524} (\bibinfo {year} {1965})}\BibitemShut
  {NoStop}%
\bibitem [{\citenamefont {Luttinger}(1966)}]{Luttinger_1966}%
  \BibitemOpen
  \bibfield  {author} {\bibinfo {author} {\bibfnamefont {J.~M.}\ \bibnamefont
  {Luttinger}},\ }\href {https://doi.org/10.1103/PhysRev.150.202} {\bibfield
  {journal} {\bibinfo  {journal} {Phys. Rev.}\ }\textbf {\bibinfo {volume}
  {150}},\ \bibinfo {pages} {202} (\bibinfo {year} {1966})}\BibitemShut
  {NoStop}%
\bibitem [{\citenamefont {Fay}\ and\ \citenamefont
  {Layzer}(1968)}]{Layzer_1968}%
  \BibitemOpen
  \bibfield  {author} {\bibinfo {author} {\bibfnamefont {D.}~\bibnamefont
  {Fay}}\ and\ \bibinfo {author} {\bibfnamefont {A.}~\bibnamefont {Layzer}},\
  }\href {https://doi.org/10.1103/PhysRevLett.20.187} {\bibfield  {journal}
  {\bibinfo  {journal} {Phys. Rev. Lett.}\ }\textbf {\bibinfo {volume} {20}},\
  \bibinfo {pages} {187} (\bibinfo {year} {1968})}\BibitemShut {NoStop}%
\bibitem [{\citenamefont {Chubukov}\ and\ \citenamefont
  {Kagan}(1989)}]{Chubukov_1989}%
  \BibitemOpen
  \bibfield  {author} {\bibinfo {author} {\bibfnamefont {A.~V.}\ \bibnamefont
  {Chubukov}}\ and\ \bibinfo {author} {\bibfnamefont {M.~Y.}\ \bibnamefont
  {Kagan}},\ }\href {https://doi.org/10.1088/0953-8984/1/19/007} {\bibfield
  {journal} {\bibinfo  {journal} {Journal of Physics: Condensed Matter}\
  }\textbf {\bibinfo {volume} {1}},\ \bibinfo {pages} {3135} (\bibinfo {year}
  {1989})}\BibitemShut {NoStop}%
\bibitem [{\citenamefont {Chubukov}(1993)}]{Chubukov_1993}%
  \BibitemOpen
  \bibfield  {author} {\bibinfo {author} {\bibfnamefont {A.~V.}\ \bibnamefont
  {Chubukov}},\ }\href {https://doi.org/10.1103/PhysRevB.48.1097} {\bibfield
  {journal} {\bibinfo  {journal} {Phys. Rev. B}\ }\textbf {\bibinfo {volume}
  {48}},\ \bibinfo {pages} {1097} (\bibinfo {year} {1993})}\BibitemShut
  {NoStop}%
\bibitem [{\citenamefont {Alexandrov}\ and\ \citenamefont
  {Kabanov}(2011)}]{Kabanov_2011}%
  \BibitemOpen
  \bibfield  {author} {\bibinfo {author} {\bibfnamefont {A.~S.}\ \bibnamefont
  {Alexandrov}}\ and\ \bibinfo {author} {\bibfnamefont {V.~V.}\ \bibnamefont
  {Kabanov}},\ }\href {https://doi.org/10.1103/PhysRevLett.106.136403}
  {\bibfield  {journal} {\bibinfo  {journal} {Phys. Rev. Lett.}\ }\textbf
  {\bibinfo {volume} {106}},\ \bibinfo {pages} {136403} (\bibinfo {year}
  {2011})}\BibitemShut {NoStop}%
\bibitem [{\citenamefont {Raghu}\ \emph {et~al.}(2012)\citenamefont {Raghu},
  \citenamefont {Berg}, \citenamefont {Chubukov},\ and\ \citenamefont
  {Kivelson}}]{Kivelson_2012}%
  \BibitemOpen
  \bibfield  {author} {\bibinfo {author} {\bibfnamefont {S.}~\bibnamefont
  {Raghu}}, \bibinfo {author} {\bibfnamefont {E.}~\bibnamefont {Berg}},
  \bibinfo {author} {\bibfnamefont {A.~V.}\ \bibnamefont {Chubukov}},\ and\
  \bibinfo {author} {\bibfnamefont {S.~A.}\ \bibnamefont {Kivelson}},\ }\href
  {https://doi.org/10.1103/PhysRevB.85.024516} {\bibfield  {journal} {\bibinfo
  {journal} {Phys. Rev. B}\ }\textbf {\bibinfo {volume} {85}},\ \bibinfo
  {pages} {024516} (\bibinfo {year} {2012})}\BibitemShut {NoStop}%
\bibitem [{\citenamefont {Friedel}(1954)}]{Friedel_1954}%
  \BibitemOpen
  \bibfield  {author} {\bibinfo {author} {\bibfnamefont {J.}~\bibnamefont
  {Friedel}},\ }\href {https://doi.org/10.1080/00018735400101233} {\bibfield
  {journal} {\bibinfo  {journal} {Advances in Physics}\ }\textbf {\bibinfo
  {volume} {3}},\ \bibinfo {pages} {446} (\bibinfo {year} {1954})},\ \Eprint
  {https://arxiv.org/abs/https://doi.org/10.1080/00018735400101233}
  {https://doi.org/10.1080/00018735400101233} \BibitemShut {NoStop}%
\bibitem [{\citenamefont {Nandkishore}\ \emph {et~al.}(2014)\citenamefont
  {Nandkishore}, \citenamefont {Thomale},\ and\ \citenamefont
  {Chubukov}}]{Rahul_2014}%
  \BibitemOpen
  \bibfield  {author} {\bibinfo {author} {\bibfnamefont {R.}~\bibnamefont
  {Nandkishore}}, \bibinfo {author} {\bibfnamefont {R.}~\bibnamefont
  {Thomale}},\ and\ \bibinfo {author} {\bibfnamefont {A.~V.}\ \bibnamefont
  {Chubukov}},\ }\href {https://doi.org/10.1103/PhysRevB.89.144501} {\bibfield
  {journal} {\bibinfo  {journal} {Phys. Rev. B}\ }\textbf {\bibinfo {volume}
  {89}},\ \bibinfo {pages} {144501} (\bibinfo {year} {2014})}\BibitemShut
  {NoStop}%
\bibitem [{\citenamefont {Kiesel}\ and\ \citenamefont
  {Thomale}(2012)}]{Thomale_2012}%
  \BibitemOpen
  \bibfield  {author} {\bibinfo {author} {\bibfnamefont {M.~L.}\ \bibnamefont
  {Kiesel}}\ and\ \bibinfo {author} {\bibfnamefont {R.}~\bibnamefont
  {Thomale}},\ }\href {https://doi.org/10.1103/PhysRevB.86.121105} {\bibfield
  {journal} {\bibinfo  {journal} {Phys. Rev. B}\ }\textbf {\bibinfo {volume}
  {86}},\ \bibinfo {pages} {121105} (\bibinfo {year} {2012})}\BibitemShut
  {NoStop}%
\bibitem [{\citenamefont {Kiesel}\ \emph {et~al.}(2013)\citenamefont {Kiesel},
  \citenamefont {Platt},\ and\ \citenamefont {Thomale}}]{Thomale_2013}%
  \BibitemOpen
  \bibfield  {author} {\bibinfo {author} {\bibfnamefont {M.~L.}\ \bibnamefont
  {Kiesel}}, \bibinfo {author} {\bibfnamefont {C.}~\bibnamefont {Platt}},\ and\
  \bibinfo {author} {\bibfnamefont {R.}~\bibnamefont {Thomale}},\ }\href
  {https://doi.org/10.1103/PhysRevLett.110.126405} {\bibfield  {journal}
  {\bibinfo  {journal} {Phys. Rev. Lett.}\ }\textbf {\bibinfo {volume} {110}},\
  \bibinfo {pages} {126405} (\bibinfo {year} {2013})}\BibitemShut {NoStop}%
\bibitem [{\citenamefont {Raghu}\ \emph {et~al.}(2010)\citenamefont {Raghu},
  \citenamefont {Kivelson},\ and\ \citenamefont {Scalapino}}]{Raghu_2010}%
  \BibitemOpen
  \bibfield  {author} {\bibinfo {author} {\bibfnamefont {S.}~\bibnamefont
  {Raghu}}, \bibinfo {author} {\bibfnamefont {S.~A.}\ \bibnamefont
  {Kivelson}},\ and\ \bibinfo {author} {\bibfnamefont {D.~J.}\ \bibnamefont
  {Scalapino}},\ }\href {https://doi.org/10.1103/PhysRevB.81.224505} {\bibfield
   {journal} {\bibinfo  {journal} {Phys. Rev. B}\ }\textbf {\bibinfo {volume}
  {81}},\ \bibinfo {pages} {224505} (\bibinfo {year} {2010})}\BibitemShut
  {NoStop}%
\bibitem [{\citenamefont {Vafek}\ \emph {et~al.}(2014)\citenamefont {Vafek},
  \citenamefont {Murray},\ and\ \citenamefont {Cvetkovic}}]{Vladimir_2014}%
  \BibitemOpen
  \bibfield  {author} {\bibinfo {author} {\bibfnamefont {O.}~\bibnamefont
  {Vafek}}, \bibinfo {author} {\bibfnamefont {J.~M.}\ \bibnamefont {Murray}},\
  and\ \bibinfo {author} {\bibfnamefont {V.}~\bibnamefont {Cvetkovic}},\ }\href
  {https://doi.org/10.1103/PhysRevLett.112.147002} {\bibfield  {journal}
  {\bibinfo  {journal} {Phys. Rev. Lett.}\ }\textbf {\bibinfo {volume} {112}},\
  \bibinfo {pages} {147002} (\bibinfo {year} {2014})}\BibitemShut {NoStop}%
\bibitem [{\citenamefont {Murray}\ and\ \citenamefont
  {Vafek}(2014)}]{Oskar_2014}%
  \BibitemOpen
  \bibfield  {author} {\bibinfo {author} {\bibfnamefont {J.~M.}\ \bibnamefont
  {Murray}}\ and\ \bibinfo {author} {\bibfnamefont {O.}~\bibnamefont {Vafek}},\
  }\href {https://doi.org/10.1103/PhysRevB.89.205119} {\bibfield  {journal}
  {\bibinfo  {journal} {Phys. Rev. B}\ }\textbf {\bibinfo {volume} {89}},\
  \bibinfo {pages} {205119} (\bibinfo {year} {2014})}\BibitemShut {NoStop}%
\bibitem [{\citenamefont {Guinea}\ and\ \citenamefont
  {Uchoa}(2012)}]{Bruno_2012}%
  \BibitemOpen
  \bibfield  {author} {\bibinfo {author} {\bibfnamefont {F.}~\bibnamefont
  {Guinea}}\ and\ \bibinfo {author} {\bibfnamefont {B.}~\bibnamefont {Uchoa}},\
  }\href {https://doi.org/10.1103/PhysRevB.86.134521} {\bibfield  {journal}
  {\bibinfo  {journal} {Phys. Rev. B}\ }\textbf {\bibinfo {volume} {86}},\
  \bibinfo {pages} {134521} (\bibinfo {year} {2012})}\BibitemShut {NoStop}%
\bibitem [{\citenamefont {Rold\'an}\ \emph {et~al.}(2013)\citenamefont
  {Rold\'an}, \citenamefont {Cappelluti},\ and\ \citenamefont
  {Guinea}}]{Guinea_2013}%
  \BibitemOpen
  \bibfield  {author} {\bibinfo {author} {\bibfnamefont {R.}~\bibnamefont
  {Rold\'an}}, \bibinfo {author} {\bibfnamefont {E.}~\bibnamefont
  {Cappelluti}},\ and\ \bibinfo {author} {\bibfnamefont {F.}~\bibnamefont
  {Guinea}},\ }\href {https://doi.org/10.1103/PhysRevB.88.054515} {\bibfield
  {journal} {\bibinfo  {journal} {Phys. Rev. B}\ }\textbf {\bibinfo {volume}
  {88}},\ \bibinfo {pages} {054515} (\bibinfo {year} {2013})}\BibitemShut
  {NoStop}%
\bibitem [{\citenamefont {Wolf}\ \emph {et~al.}(2022)\citenamefont {Wolf},
  \citenamefont {Di~Sante}, \citenamefont {Schwemmer}, \citenamefont
  {Thomale},\ and\ \citenamefont {Rachel}}]{Stephan_2022}%
  \BibitemOpen
  \bibfield  {author} {\bibinfo {author} {\bibfnamefont {S.}~\bibnamefont
  {Wolf}}, \bibinfo {author} {\bibfnamefont {D.}~\bibnamefont {Di~Sante}},
  \bibinfo {author} {\bibfnamefont {T.}~\bibnamefont {Schwemmer}}, \bibinfo
  {author} {\bibfnamefont {R.}~\bibnamefont {Thomale}},\ and\ \bibinfo {author}
  {\bibfnamefont {S.}~\bibnamefont {Rachel}},\ }\href
  {https://doi.org/10.1103/PhysRevLett.128.167002} {\bibfield  {journal}
  {\bibinfo  {journal} {Phys. Rev. Lett.}\ }\textbf {\bibinfo {volume} {128}},\
  \bibinfo {pages} {167002} (\bibinfo {year} {2022})}\BibitemShut {NoStop}%
\bibitem [{\citenamefont {Lin}\ and\ \citenamefont
  {Nandkishore}(2018)}]{Rahul_2018}%
  \BibitemOpen
  \bibfield  {author} {\bibinfo {author} {\bibfnamefont {Y.-P.}\ \bibnamefont
  {Lin}}\ and\ \bibinfo {author} {\bibfnamefont {R.~M.}\ \bibnamefont
  {Nandkishore}},\ }\href {https://doi.org/10.1103/PhysRevB.98.214521}
  {\bibfield  {journal} {\bibinfo  {journal} {Phys. Rev. B}\ }\textbf {\bibinfo
  {volume} {98}},\ \bibinfo {pages} {214521} (\bibinfo {year}
  {2018})}\BibitemShut {NoStop}%
\bibitem [{\citenamefont {Dürrnagel}\ \emph {et~al.}(2022)\citenamefont
  {Dürrnagel}, \citenamefont {Beyer}, \citenamefont {Thomale},\ and\
  \citenamefont {Schwemmer}}]{Schwemmer_2022}%
  \BibitemOpen
  \bibfield  {author} {\bibinfo {author} {\bibfnamefont {M.}~\bibnamefont
  {Dürrnagel}}, \bibinfo {author} {\bibfnamefont {J.}~\bibnamefont {Beyer}},
  \bibinfo {author} {\bibfnamefont {R.}~\bibnamefont {Thomale}},\ and\ \bibinfo
  {author} {\bibfnamefont {T.}~\bibnamefont {Schwemmer}},\ }\href
  {https://doi.org/10.1140/epjb/s10051-022-00371-4} {\bibfield  {journal}
  {\bibinfo  {journal} {The European Physical Journal B}\ }\textbf {\bibinfo
  {volume} {95}},\ \bibinfo {pages} {112} (\bibinfo {year} {2022})}\BibitemShut
  {NoStop}%
\bibitem [{\citenamefont {Wolf}\ \emph {et~al.}(2018)\citenamefont {Wolf},
  \citenamefont {Schmidt},\ and\ \citenamefont {Rachel}}]{Wolf_2018}%
  \BibitemOpen
  \bibfield  {author} {\bibinfo {author} {\bibfnamefont {S.}~\bibnamefont
  {Wolf}}, \bibinfo {author} {\bibfnamefont {T.~L.}\ \bibnamefont {Schmidt}},\
  and\ \bibinfo {author} {\bibfnamefont {S.}~\bibnamefont {Rachel}},\ }\href
  {https://doi.org/10.1103/PhysRevB.98.174515} {\bibfield  {journal} {\bibinfo
  {journal} {Phys. Rev. B}\ }\textbf {\bibinfo {volume} {98}},\ \bibinfo
  {pages} {174515} (\bibinfo {year} {2018})}\BibitemShut {NoStop}%
\bibitem [{\citenamefont {Nandkishore}\ and\ \citenamefont
  {Sondhi}(2017)}]{Sondhi_2017}%
  \BibitemOpen
  \bibfield  {author} {\bibinfo {author} {\bibfnamefont {R.~M.}\ \bibnamefont
  {Nandkishore}}\ and\ \bibinfo {author} {\bibfnamefont {S.~L.}\ \bibnamefont
  {Sondhi}},\ }\href {https://doi.org/10.1103/PhysRevX.7.041021} {\bibfield
  {journal} {\bibinfo  {journal} {Phys. Rev. X}\ }\textbf {\bibinfo {volume}
  {7}},\ \bibinfo {pages} {041021} (\bibinfo {year} {2017})}\BibitemShut
  {NoStop}%
\bibitem [{\citenamefont {Kagan}\ \emph
  {et~al.}(2016{\natexlab{a}})\citenamefont {Kagan}, \citenamefont {Mitskan},\
  and\ \citenamefont {Korovushkin}}]{Korovushkin_2016}%
  \BibitemOpen
  \bibfield  {author} {\bibinfo {author} {\bibfnamefont {M.~Y.}\ \bibnamefont
  {Kagan}}, \bibinfo {author} {\bibfnamefont {V.~A.}\ \bibnamefont {Mitskan}},\
  and\ \bibinfo {author} {\bibfnamefont {M.~M.}\ \bibnamefont {Korovushkin}},\
  }\href {https://doi.org/10.1007/s10948-016-3384-7} {\bibfield  {journal}
  {\bibinfo  {journal} {Journal of Superconductivity and Novel Magnetism}\
  }\textbf {\bibinfo {volume} {29}},\ \bibinfo {pages} {1043} (\bibinfo {year}
  {2016}{\natexlab{a}})}\BibitemShut {NoStop}%
\bibitem [{\citenamefont {Kagan}\ \emph
  {et~al.}(2016{\natexlab{b}})\citenamefont {Kagan}, \citenamefont {Mitskan},\
  and\ \citenamefont {Korovushkin}}]{Korovushkin_2016_1}%
  \BibitemOpen
  \bibfield  {author} {\bibinfo {author} {\bibfnamefont {M.~Y.}\ \bibnamefont
  {Kagan}}, \bibinfo {author} {\bibfnamefont {V.~A.}\ \bibnamefont {Mitskan}},\
  and\ \bibinfo {author} {\bibfnamefont {M.~M.}\ \bibnamefont {Korovushkin}},\
  }\href {https://doi.org/10.1007/s10909-015-1427-2} {\bibfield  {journal}
  {\bibinfo  {journal} {Journal of Low Temperature Physics}\ }\textbf {\bibinfo
  {volume} {185}},\ \bibinfo {pages} {508} (\bibinfo {year}
  {2016}{\natexlab{b}})}\BibitemShut {NoStop}%
\bibitem [{\citenamefont {Kagan}\ \emph
  {et~al.}(2015{\natexlab{a}})\citenamefont {Kagan}, \citenamefont {Mitskan},\
  and\ \citenamefont {Korovushkin}}]{Korovushkin_2015}%
  \BibitemOpen
  \bibfield  {author} {\bibinfo {author} {\bibfnamefont {M.~Y.}\ \bibnamefont
  {Kagan}}, \bibinfo {author} {\bibfnamefont {V.~A.}\ \bibnamefont {Mitskan}},\
  and\ \bibinfo {author} {\bibfnamefont {M.~M.}\ \bibnamefont {Korovushkin}},\
  }\href {https://doi.org/10.3367/UFNe.0185.201508a.0785} {\bibfield  {journal}
  {\bibinfo  {journal} {Physics-Uspekhi}\ }\textbf {\bibinfo {volume} {58}},\
  \bibinfo {pages} {733} (\bibinfo {year} {2015}{\natexlab{a}})}\BibitemShut
  {NoStop}%
\bibitem [{\citenamefont {Kagan}\ \emph
  {et~al.}(2015{\natexlab{b}})\citenamefont {Kagan}, \citenamefont {Mitskan},\
  and\ \citenamefont {Korovushkin}}]{Korovushkin_2015_1}%
  \BibitemOpen
  \bibfield  {author} {\bibinfo {author} {\bibfnamefont {M.~Y.}\ \bibnamefont
  {Kagan}}, \bibinfo {author} {\bibfnamefont {V.~A.}\ \bibnamefont {Mitskan}},\
  and\ \bibinfo {author} {\bibfnamefont {M.~M.}\ \bibnamefont {Korovushkin}},\
  }\href {https://doi.org/10.1140/epjb/e2015-60198-x} {\bibfield  {journal}
  {\bibinfo  {journal} {The European Physical Journal B}\ }\textbf {\bibinfo
  {volume} {88}},\ \bibinfo {pages} {157} (\bibinfo {year}
  {2015}{\natexlab{b}})}\BibitemShut {NoStop}%
\bibitem [{\citenamefont {Gonz\'alez}\ and\ \citenamefont
  {Stauber}(2019)}]{Stauber_2019}%
  \BibitemOpen
  \bibfield  {author} {\bibinfo {author} {\bibfnamefont {J.}~\bibnamefont
  {Gonz\'alez}}\ and\ \bibinfo {author} {\bibfnamefont {T.}~\bibnamefont
  {Stauber}},\ }\href {https://doi.org/10.1103/PhysRevLett.122.026801}
  {\bibfield  {journal} {\bibinfo  {journal} {Phys. Rev. Lett.}\ }\textbf
  {\bibinfo {volume} {122}},\ \bibinfo {pages} {026801} (\bibinfo {year}
  {2019})}\BibitemShut {NoStop}%
\bibitem [{\citenamefont {Cho}\ \emph {et~al.}(2013)\citenamefont {Cho},
  \citenamefont {Thomale}, \citenamefont {Raghu},\ and\ \citenamefont
  {Kivelson}}]{Srinivas_2013}%
  \BibitemOpen
  \bibfield  {author} {\bibinfo {author} {\bibfnamefont {W.}~\bibnamefont
  {Cho}}, \bibinfo {author} {\bibfnamefont {R.}~\bibnamefont {Thomale}},
  \bibinfo {author} {\bibfnamefont {S.}~\bibnamefont {Raghu}},\ and\ \bibinfo
  {author} {\bibfnamefont {S.~A.}\ \bibnamefont {Kivelson}},\ }\href
  {https://doi.org/10.1103/PhysRevB.88.064505} {\bibfield  {journal} {\bibinfo
  {journal} {Phys. Rev. B}\ }\textbf {\bibinfo {volume} {88}},\ \bibinfo
  {pages} {064505} (\bibinfo {year} {2013})}\BibitemShut {NoStop}%
\bibitem [{\citenamefont {Jiang}\ \emph {et~al.}(2015)\citenamefont {Jiang},
  \citenamefont {Tang}, \citenamefont {Pan}, \citenamefont {Liu}, \citenamefont
  {Niu}, \citenamefont {Wang}, \citenamefont {Xu}, \citenamefont {Yang},
  \citenamefont {Xie}, \citenamefont {Song}, \citenamefont {Dudin},
  \citenamefont {Kim}, \citenamefont {Hoesch}, \citenamefont {Das},
  \citenamefont {Vobornik}, \citenamefont {Wan},\ and\ \citenamefont
  {Feng}}]{Jiang_2015}%
  \BibitemOpen
  \bibfield  {author} {\bibinfo {author} {\bibfnamefont {J.}~\bibnamefont
  {Jiang}}, \bibinfo {author} {\bibfnamefont {F.}~\bibnamefont {Tang}},
  \bibinfo {author} {\bibfnamefont {X.~C.}\ \bibnamefont {Pan}}, \bibinfo
  {author} {\bibfnamefont {H.~M.}\ \bibnamefont {Liu}}, \bibinfo {author}
  {\bibfnamefont {X.~H.}\ \bibnamefont {Niu}}, \bibinfo {author} {\bibfnamefont
  {Y.~X.}\ \bibnamefont {Wang}}, \bibinfo {author} {\bibfnamefont {D.~F.}\
  \bibnamefont {Xu}}, \bibinfo {author} {\bibfnamefont {H.~F.}\ \bibnamefont
  {Yang}}, \bibinfo {author} {\bibfnamefont {B.~P.}\ \bibnamefont {Xie}},
  \bibinfo {author} {\bibfnamefont {F.~Q.}\ \bibnamefont {Song}}, \bibinfo
  {author} {\bibfnamefont {P.}~\bibnamefont {Dudin}}, \bibinfo {author}
  {\bibfnamefont {T.~K.}\ \bibnamefont {Kim}}, \bibinfo {author} {\bibfnamefont
  {M.}~\bibnamefont {Hoesch}}, \bibinfo {author} {\bibfnamefont {P.~K.}\
  \bibnamefont {Das}}, \bibinfo {author} {\bibfnamefont {I.}~\bibnamefont
  {Vobornik}}, \bibinfo {author} {\bibfnamefont {X.~G.}\ \bibnamefont {Wan}},\
  and\ \bibinfo {author} {\bibfnamefont {D.~L.}\ \bibnamefont {Feng}},\ }\href
  {https://doi.org/10.1103/PhysRevLett.115.166601} {\bibfield  {journal}
  {\bibinfo  {journal} {Phys. Rev. Lett.}\ }\textbf {\bibinfo {volume} {115}},\
  \bibinfo {pages} {166601} (\bibinfo {year} {2015})}\BibitemShut {NoStop}%
\bibitem [{\citenamefont {Qian}\ \emph {et~al.}(2014)\citenamefont {Qian},
  \citenamefont {Liu}, \citenamefont {Fu},\ and\ \citenamefont
  {Li}}]{Qian_2014}%
  \BibitemOpen
  \bibfield  {author} {\bibinfo {author} {\bibfnamefont {X.}~\bibnamefont
  {Qian}}, \bibinfo {author} {\bibfnamefont {J.}~\bibnamefont {Liu}}, \bibinfo
  {author} {\bibfnamefont {L.}~\bibnamefont {Fu}},\ and\ \bibinfo {author}
  {\bibfnamefont {J.}~\bibnamefont {Li}},\ }\href
  {https://doi.org/10.1126/science.1256815} {\bibfield  {journal} {\bibinfo
  {journal} {Science}\ }\textbf {\bibinfo {volume} {346}},\ \bibinfo {pages}
  {1344} (\bibinfo {year} {2014})},\ \Eprint
  {https://arxiv.org/abs/https://www.science.org/doi/pdf/10.1126/science.1256815}
  {https://www.science.org/doi/pdf/10.1126/science.1256815} \BibitemShut
  {NoStop}%
\bibitem [{\citenamefont {Ok}\ \emph {et~al.}(2019)\citenamefont {Ok},
  \citenamefont {Muechler}, \citenamefont {Di~Sante}, \citenamefont
  {Sangiovanni}, \citenamefont {Thomale},\ and\ \citenamefont
  {Neupert}}]{Ok_2019}%
  \BibitemOpen
  \bibfield  {author} {\bibinfo {author} {\bibfnamefont {S.}~\bibnamefont
  {Ok}}, \bibinfo {author} {\bibfnamefont {L.}~\bibnamefont {Muechler}},
  \bibinfo {author} {\bibfnamefont {D.}~\bibnamefont {Di~Sante}}, \bibinfo
  {author} {\bibfnamefont {G.}~\bibnamefont {Sangiovanni}}, \bibinfo {author}
  {\bibfnamefont {R.}~\bibnamefont {Thomale}},\ and\ \bibinfo {author}
  {\bibfnamefont {T.}~\bibnamefont {Neupert}},\ }\href
  {https://doi.org/10.1103/PhysRevB.99.121105} {\bibfield  {journal} {\bibinfo
  {journal} {Physical Review B}\ }\textbf {\bibinfo {volume} {99}},\ \bibinfo
  {pages} {121105} (\bibinfo {year} {2019})}\BibitemShut {NoStop}%
\bibitem [{\citenamefont {Choe}\ \emph {et~al.}(2016)\citenamefont {Choe},
  \citenamefont {Sung},\ and\ \citenamefont {Chang}}]{Choe_2016}%
  \BibitemOpen
  \bibfield  {author} {\bibinfo {author} {\bibfnamefont {D.-H.}\ \bibnamefont
  {Choe}}, \bibinfo {author} {\bibfnamefont {H.-J.}\ \bibnamefont {Sung}},\
  and\ \bibinfo {author} {\bibfnamefont {K.~J.}\ \bibnamefont {Chang}},\ }\href
  {https://doi.org/10.1103/PhysRevB.93.125109} {\bibfield  {journal} {\bibinfo
  {journal} {Phys. Rev. B}\ }\textbf {\bibinfo {volume} {93}},\ \bibinfo
  {pages} {125109} (\bibinfo {year} {2016})}\BibitemShut {NoStop}%
\bibitem [{\citenamefont {Zheng}\ \emph {et~al.}(2016)\citenamefont {Zheng},
  \citenamefont {Cai}, \citenamefont {Ge}, \citenamefont {Zhang}, \citenamefont
  {Liu}, \citenamefont {Lu}, \citenamefont {Zhang}, \citenamefont {Qiu},
  \citenamefont {Taniguchi}, \citenamefont {Watanabe}, \citenamefont {Jia},
  \citenamefont {Qi}, \citenamefont {Chen}, \citenamefont {Sun},\ and\
  \citenamefont {Feng}}]{Zheng_2016}%
  \BibitemOpen
  \bibfield  {author} {\bibinfo {author} {\bibfnamefont {F.}~\bibnamefont
  {Zheng}}, \bibinfo {author} {\bibfnamefont {C.}~\bibnamefont {Cai}}, \bibinfo
  {author} {\bibfnamefont {S.}~\bibnamefont {Ge}}, \bibinfo {author}
  {\bibfnamefont {X.}~\bibnamefont {Zhang}}, \bibinfo {author} {\bibfnamefont
  {X.}~\bibnamefont {Liu}}, \bibinfo {author} {\bibfnamefont {H.}~\bibnamefont
  {Lu}}, \bibinfo {author} {\bibfnamefont {Y.}~\bibnamefont {Zhang}}, \bibinfo
  {author} {\bibfnamefont {J.}~\bibnamefont {Qiu}}, \bibinfo {author}
  {\bibfnamefont {T.}~\bibnamefont {Taniguchi}}, \bibinfo {author}
  {\bibfnamefont {K.}~\bibnamefont {Watanabe}}, \bibinfo {author}
  {\bibfnamefont {S.}~\bibnamefont {Jia}}, \bibinfo {author} {\bibfnamefont
  {J.}~\bibnamefont {Qi}}, \bibinfo {author} {\bibfnamefont {J.-H.}\
  \bibnamefont {Chen}}, \bibinfo {author} {\bibfnamefont {D.}~\bibnamefont
  {Sun}},\ and\ \bibinfo {author} {\bibfnamefont {J.}~\bibnamefont {Feng}},\
  }\href {https://doi.org/https://doi.org/10.1002/adma.201600100} {\bibfield
  {journal} {\bibinfo  {journal} {Advanced Materials}\ }\textbf {\bibinfo
  {volume} {28}},\ \bibinfo {pages} {4845} (\bibinfo {year} {2016})},\ \Eprint
  {https://arxiv.org/abs/https://onlinelibrary.wiley.com/doi/pdf/10.1002/adma.201600100}
  {https://onlinelibrary.wiley.com/doi/pdf/10.1002/adma.201600100} \BibitemShut
  {NoStop}%
\bibitem [{\citenamefont {Zhao}\ \emph {et~al.}(2021)\citenamefont {Zhao},
  \citenamefont {Runburg}, \citenamefont {Fei}, \citenamefont {Mutch},
  \citenamefont {Malinowski}, \citenamefont {Sun}, \citenamefont {Huang},
  \citenamefont {Pesin}, \citenamefont {Cui}, \citenamefont {Xu}, \citenamefont
  {Chu},\ and\ \citenamefont {Cobden}}]{zhao_2021}%
  \BibitemOpen
  \bibfield  {author} {\bibinfo {author} {\bibfnamefont {W.}~\bibnamefont
  {Zhao}}, \bibinfo {author} {\bibfnamefont {E.}~\bibnamefont {Runburg}},
  \bibinfo {author} {\bibfnamefont {Z.}~\bibnamefont {Fei}}, \bibinfo {author}
  {\bibfnamefont {J.}~\bibnamefont {Mutch}}, \bibinfo {author} {\bibfnamefont
  {P.}~\bibnamefont {Malinowski}}, \bibinfo {author} {\bibfnamefont
  {B.}~\bibnamefont {Sun}}, \bibinfo {author} {\bibfnamefont {X.}~\bibnamefont
  {Huang}}, \bibinfo {author} {\bibfnamefont {D.}~\bibnamefont {Pesin}},
  \bibinfo {author} {\bibfnamefont {Y.-T.}\ \bibnamefont {Cui}}, \bibinfo
  {author} {\bibfnamefont {X.}~\bibnamefont {Xu}}, \bibinfo {author}
  {\bibfnamefont {J.-H.}\ \bibnamefont {Chu}},\ and\ \bibinfo {author}
  {\bibfnamefont {D.~H.}\ \bibnamefont {Cobden}},\ }\href
  {https://doi.org/10.1103/PhysRevX.11.041034} {\bibfield  {journal} {\bibinfo
  {journal} {Phys. Rev. X}\ }\textbf {\bibinfo {volume} {11}},\ \bibinfo
  {pages} {041034} (\bibinfo {year} {2021})}\BibitemShut {NoStop}%
\bibitem [{\citenamefont {Lapa}\ \emph {et~al.}(2021)\citenamefont {Lapa},
  \citenamefont {Cheng},\ and\ \citenamefont {Wang}}]{Lapa_2021}%
  \BibitemOpen
  \bibfield  {author} {\bibinfo {author} {\bibfnamefont {M.~F.}\ \bibnamefont
  {Lapa}}, \bibinfo {author} {\bibfnamefont {M.}~\bibnamefont {Cheng}},\ and\
  \bibinfo {author} {\bibfnamefont {Y.}~\bibnamefont {Wang}},\ }\href
  {https://doi.org/10.21468/SciPostPhys.11.5.086} {\bibfield  {journal}
  {\bibinfo  {journal} {SciPost Phys.}\ }\textbf {\bibinfo {volume} {11}},\
  \bibinfo {pages} {86} (\bibinfo {year} {2021})}\BibitemShut {NoStop}%
\bibitem [{\citenamefont {Muechler}\ \emph {et~al.}(2016)\citenamefont
  {Muechler}, \citenamefont {Alexandradinata}, \citenamefont {Neupert},\ and\
  \citenamefont {Car}}]{Muechler_2016}%
  \BibitemOpen
  \bibfield  {author} {\bibinfo {author} {\bibfnamefont {L.}~\bibnamefont
  {Muechler}}, \bibinfo {author} {\bibfnamefont {A.}~\bibnamefont
  {Alexandradinata}}, \bibinfo {author} {\bibfnamefont {T.}~\bibnamefont
  {Neupert}},\ and\ \bibinfo {author} {\bibfnamefont {R.}~\bibnamefont {Car}},\
  }\href {https://doi.org/10.1103/PhysRevX.6.041069} {\bibfield  {journal}
  {\bibinfo  {journal} {Phys. Rev. X}\ }\textbf {\bibinfo {volume} {6}},\
  \bibinfo {pages} {041069} (\bibinfo {year} {2016})}\BibitemShut {NoStop}%
\bibitem [{\citenamefont {Asaba}\ \emph {et~al.}(2018)\citenamefont {Asaba},
  \citenamefont {Wang}, \citenamefont {Li}, \citenamefont {Xiang},
  \citenamefont {Tinsman}, \citenamefont {Chen}, \citenamefont {Zhou},
  \citenamefont {Zhao}, \citenamefont {Laleyan}, \citenamefont {Li},
  \citenamefont {Mi},\ and\ \citenamefont {Li}}]{Li_2018}%
  \BibitemOpen
  \bibfield  {author} {\bibinfo {author} {\bibfnamefont {T.}~\bibnamefont
  {Asaba}}, \bibinfo {author} {\bibfnamefont {Y.}~\bibnamefont {Wang}},
  \bibinfo {author} {\bibfnamefont {G.}~\bibnamefont {Li}}, \bibinfo {author}
  {\bibfnamefont {Z.}~\bibnamefont {Xiang}}, \bibinfo {author} {\bibfnamefont
  {C.}~\bibnamefont {Tinsman}}, \bibinfo {author} {\bibfnamefont
  {L.}~\bibnamefont {Chen}}, \bibinfo {author} {\bibfnamefont {S.}~\bibnamefont
  {Zhou}}, \bibinfo {author} {\bibfnamefont {S.}~\bibnamefont {Zhao}}, \bibinfo
  {author} {\bibfnamefont {D.}~\bibnamefont {Laleyan}}, \bibinfo {author}
  {\bibfnamefont {Y.}~\bibnamefont {Li}}, \bibinfo {author} {\bibfnamefont
  {Z.}~\bibnamefont {Mi}},\ and\ \bibinfo {author} {\bibfnamefont
  {L.}~\bibnamefont {Li}},\ }\href {https://doi.org/10.1038/s41598-018-24736-x}
  {\bibfield  {journal} {\bibinfo  {journal} {Scientific Reports}\ }\textbf
  {\bibinfo {volume} {8}},\ \bibinfo {pages} {6520} (\bibinfo {year}
  {2018})}\BibitemShut {NoStop}%
\end{thebibliography}%
\end{document}